%% file: main_arxiv.tex
\pdfoutput=1
\documentclass[%
prx,twocolumn,
notitlepage,
superscriptaddress,
 amsmath,amssymb,
 aps,
prx,
floatfix,
longbibliography
]{revtex4-1}

\usepackage[T1]{fontenc}
\usepackage{float}
\usepackage{color}
\usepackage{amsmath}
\usepackage{amssymb}
\usepackage{bm}
\usepackage{graphicx}
\usepackage[usenames,dvipsnames]{xcolor}
\usepackage[%
  colorlinks=true,
  urlcolor=blue,
  linkcolor=blue,
  citecolor=blue
]{hyperref}

\usepackage{array}
\definecolor{bleu}{rgb}{0.16,0.2.5,0.36}

\newcommand{\V}{\raisebox{-0.20\height}{\includegraphics[height=2.5ex]{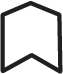}}}
\newcommand{\Square}{\raisebox{-0.20\height}{\includegraphics[height=2.5ex]{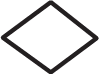}}}
\newcommand{\House}{\raisebox{-0.20\height}{\includegraphics[height=2.5ex]{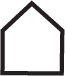}}}
\newcommand{\SI}{\raisebox{-0.20\height}{\includegraphics[height=2.5ex]{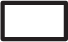}}}
\newcommand{\SB}{\raisebox{-0.20\height}{\includegraphics[height=2.5ex]{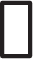}}}
\newcommand{\AB}{\raisebox{-0.20\height}{\includegraphics[height=2.5ex]{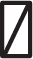}}}

\newcommand{\AI}{\raisebox{-0.20\height}{\includegraphics[height=2.5ex]{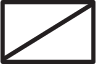}}}

\newcommand{\VV}{\raisebox{-0.20\height}{\includegraphics[height=2.5ex]{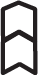}}}
\newcommand{\VS}{\raisebox{-0.20\height}{\includegraphics[height=2.5ex]{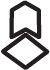}}}
\newcommand{\SV}{\raisebox{-0.20\height}{\includegraphics[height=2.5ex]{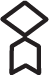}}}
\newcommand{\SSS}{\raisebox{-0.20\height}{\includegraphics[height=2.5ex]{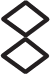}}}

\newcommand{\HS}{\raisebox{-0.20\height}{\includegraphics[height=2.5ex]{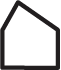}}}

\newcommand{\HA}{\raisebox{-0.20\height}{\includegraphics[height=2.5ex]{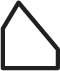}}}

\newcommand{\HT}{\raisebox{-0.20\height}{\includegraphics[height=2.5ex]{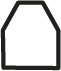}}}

\newcommand{\SD}{\raisebox{-0.20\height}{\includegraphics[height=2.5ex]{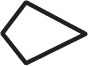}}}

\newcommand{\SR}{\raisebox{-0.20\height}{\includegraphics[height=2.5ex]{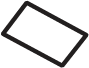}}}

\newcommand{\SH}{\raisebox{-0.20\height}{\includegraphics[height=2.5ex]{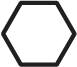}}}

\newcommand{\VT}{\raisebox{-0.20\height}{\includegraphics[height=2.5ex]{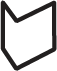}}}

\newcommand{\VA}{\raisebox{-0.20\height}{\includegraphics[height=2.5ex]{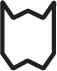}}}
\newcommand{\VU}{\raisebox{-0.20\height}{\includegraphics[height=2.5ex]{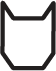}}}
\newcommand{\EDF}{\textbf{Fig.}}%to use for submission to journal
\usepackage{lipsum} % for dummy text
\begin{document}
\pdfoutput=1

% \title{Multishape mechanical self-assembly using loading history}
%\title{Multishape self-assembly kirigami with tunable rigidity}
%\title{Self-guided folding of pluripotent metamaterials}
\title{Folding of multishape metamaterials guided by plasticity}
%\title{Self-guided kirigami shape-morphing
%Multishape self-folding kirigami with tunable rigidity}
%History-dependent shape-shifting kirigami}
\author{Wenfeng Liu}
\affiliation{Institute of Physics, Universiteit van Amsterdam, 1098 XH Amsterdam, The Netherlands}
\affiliation{John A. Paulson School of Engineering and Applied Sciences, Harvard University, Cambridge, Massachusetts 02138, USA}
\author{Jean-Baptiste Chombart}
\affiliation{Institute of Physics, Universiteit van Amsterdam, 1098 XH Amsterdam, The Netherlands}
\author{Bernard Ennis}
\affiliation{VYTL Works, Amsterdam}
\author{Corentin Coulais}
\affiliation{Institute of Physics, Universiteit van Amsterdam, 1098 XH Amsterdam, The Netherlands}
\affiliation{VYTL Works, Amsterdam}

\begin{abstract}
Multishape metamaterials can morph into several different shapes, and offer exciting prospects in areas as diverse as tunable optics, wearable electronics, soft robotics, and aerospace. 
Yet, multishape metamaterials are typically actuated with complex multiaxial or multiphysics loading.
Here, we introduce a uniaxial folding strategy that turns kirigami---2D sheets patterned with cuts---into multishape metamaterials with tunable rigidity. 
Our folding approach leverages plasticity and loading history to guide multiple and tunable distinct shape changes upon cycles of stretching and compression. 
The deployed shapes exhibit a wide range of mechanical properties spanning three orders of magnitude. 
Self-guided plastic folding provides a uniquely scalable route to deploy load-bearing metamaterials from the micrometer up to the industrial scale.
\end{abstract}
\maketitle
% \linenumbers
% \switchlinenumbers
% \section{Significance}
From protein folding~\cite{tsuboyama2023mega} to organisms that flex their shapes during prey capture~\cite{flaum2024curved} and flies that spread their wings~\cite{hadjaje2024mechanics, faber2018bioinspired}, there is an abundance of shape-changing structures in the biological world. Shape morphing metamaterials~\cite{dudek2025shape} aim to emulate such shape morphing capabilities to create novel generations of optical devices~\cite{smart2024magnetically}, wearables~\cite{bai2022dynamically, xue2020mechanically}, robots~\cite{melancon2022inflatable, hwang2022shape, tang2019programmable},
buildings~\cite{melancon2021multistable, alhafnawi2026architectural}, and spacecraft mechanisms~\cite{tibert2002deployable,dold2026continuous}. However, two challenges remain largely unaddressed. How to achieve multiple morphing shapes? How to give those shapes tunable rigidity?

First, although many metamaterials can now morph into multiple shapes~\cite{dudek2025shape, misseroni2024origami,dieleman2020jigsaw, dreier2025beaded, hong2025reprogrammable, yang2023morphing, meeussen2023multistable, dykstra2024inverse, van2025prospecting}, they require complex multiaxial actuation to do so, typically through manual manipulation. Taking multishape metamaterials literally out of our hands will require a simpler single-input actuation strategy.
There are attempts to create multishape metamaterials under single-input actuation based on multistability~\cite{fu2018morphable, meng2022deployable, melancon2022inflatable, liu2024controlled, kamp2025reprogrammable, melancon2022inflatable} and on the loading speed of viscoelasticity~\cite{janbaz2024diffusive}. 
However, these metamaterials require multiple materials, and are therefore are challenging to scale up or down. 
Second, shape-morphing structures are usually monostable and return to their initial configuration once the actuation is removed~\cite{janbaz2024diffusive, siefert2019bio}. To maintain the morphed shapes with rigidity, one approach is to exploit multistability~\cite{chen2021bistable, meng2022deployable, meeussen2023multistable, liu2023snap}. Another approach is to induce plastic deformation~\cite{rafsanjani2017buckling, hwang2022shape, charpentier2024morpho, baker2018spin, massimino2026metal}. 
However, the obtained shapes exhibit either bending-dominated deformation modes~\cite{rafsanjani2017buckling, meng2022deployable, charpentier2024morpho} or stretching-dominated deformation modes~\cite{baker2018spin, massimino2026metal}, with limited rigidity and tunability.
Loading history from a single input as an effective design tool has been used for mechanical computing~\cite{liu2024controlled, bense2021complex} and soft robotics~\cite{kamp2025reprogrammable, melancon2022inflatable}, but is rarely explored in 3D shape-changing structures.

Here, we introduce loading history and plasticity into shape-changing kirigami whose 3D morphed shapes can be tuned from bending-dominated to stretching-dominated.  Thus, our shape-morphing kirigami is pluripotent~\cite{cho2014engineering,dieleman2020jigsaw} for a wide range of mechanical performance and is easy to produce and assemble from readily available materials.

\section*{Self-guided plastic folding}
\begin{figure*}[htb!]
    \centering
\includegraphics[width=1\linewidth]{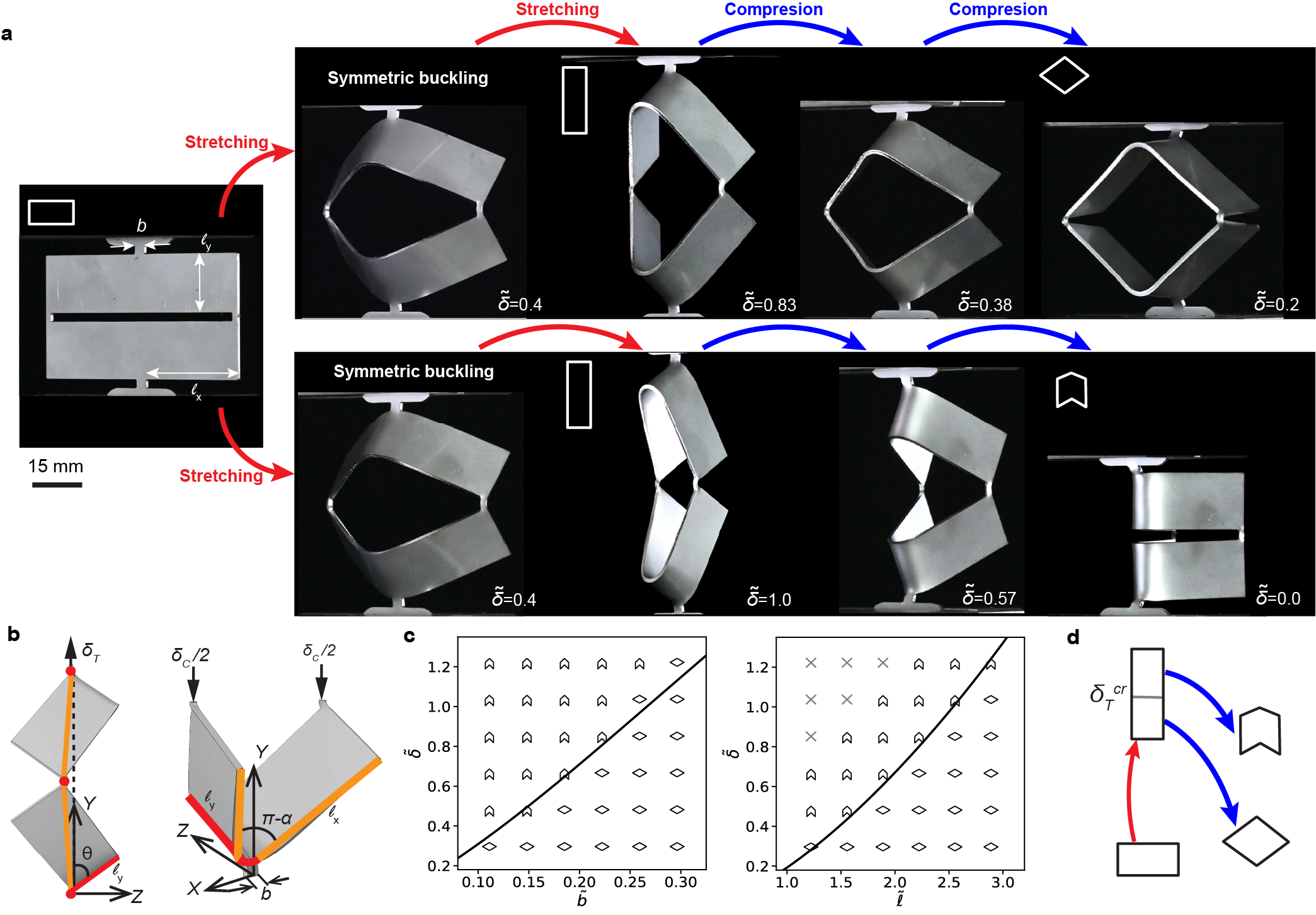}
\caption{\textbf{Folding of kirigami into multiple shapes guided by plasticity.} \textbf{a}, Snapshots of the kirigami unit cell under two different stretch strokes $\tilde{\delta}=\delta_T/2\ell_y$ before unloading back. \protect\SI\ and \protect\SB\  denote initial flat and symmetric buckling modes, respectively. \protect\Square\ and \protect\V\ denote the two folding shapes. \textbf{b} left, Kirigami unit cell under compression from the side view. Red circles indicate the rotating hinges. \textbf{b} right, Half kirigami unit cell under compression from an oblique view. The red bar represents the plate's bending hinge. \textbf{c} Stretch stroke $\tilde{\delta}$ vs. (c left) normalized hinge size $\tilde{b}=b/\ell_y$ and (c left) normalized unit width $\tilde{\ell}=\ell_x/\ell_y$. The markers are finite-element results and the gray crosses denote hinge fracture under stretching. The solid lines in the panel are the prediction of Eq.~\eqref{foding boudanry} for a given $\tilde{\ell}=1.86$ (left) and $\tilde{b}=0.17$ (right), respectively. \textbf{d}, A transition graph of the kirigami contains four states of configurations and the loading paths. A lying rectangle and a standing rectangle indicate the initial state and the symmetric buckling state, respectively. A shield and a diamond indicate the final folding state in panel a. Red (blue) arrows correspond to up (down) transitions at stretching (compression) as indicated in the stroke. $\delta_T^{cr}$ indicates the critical stretch obtained from panel c. Stretching above $\delta_T^{cr}$ induces a chevron shape under compression; otherwise, a diamond shape forms.}
\label{fig:1 folding shapes}
\end{figure*}

We first consider a kirigami unit cell with the standard parallel-cut pattern~\cite{rafsanjani2017buckling,janbaz2024diffusive, yang2018multistable, tao2023engineering} made from an elastoplastic material, which naturally buckles symmetrically under stretching (Fig.~\ref{fig:1 folding shapes}a \protect\SB\ and Methods). Secondly, we unload it back, and surprisingly, the 3D shape folds itself towards a diamond (\protect\Square) or 
a chevron (\protect\V) shape depending on the stretch value before unloading (Fig.~\ref{fig:1 folding shapes}a and Supplementary Movie 1). This is in stark contrast with the elastic case, where the kirigami would unbuckle back to a flat shape. 

Where does this counterintuitive response come from? What determines the transition between the \protect\Square\ and the \protect\V\ shapes? To answer these questions, we conduct theoretical analysis on the plastic deformation of kirigami under stretching and compression.  

We idealize the deformation of kirigami into two parts: the rotation between the two connected plates around the hinge joints (Fig.~\ref{fig:1 folding shapes}b left, red dots) and the bending within the two deformed plates around the plate joints (Fig.~\ref{fig:1 folding shapes}b right, red bar), which can be described by the joint angle $\theta$ and plate angle $\alpha$, respectively. Both $\theta$ and $\alpha$ increase with the square root of stretch $\tilde{\delta}_T$ (see Supplementary Information), 
\begin{equation}
\theta \approx \left(\frac{2}{1+2\tilde{b}\tilde{\ell}}\tilde{\delta}\right)^{1/2},
\quad
\alpha \approx  \left(\frac{8(1+\tilde{b}\tilde{\ell})^2}{\tilde{\ell}^2(1+2\tilde{b}\tilde{\ell})}\tilde{\delta}\right)^{1/2}
     \label{stretching angle}
\end{equation}
where $\tilde{b} = \frac{b}{\ell_y}$, $\tilde{\ell}= \frac{\ell_x}{\ell_y}$, and $\tilde{\delta} = \frac{\delta_T}{2\ell_y}$ (Fig.~\ref{fig:1 folding shapes}b). This square root singularity is a hallmark of buckling instabilities. 
% This theory assumes that joints act as plastic hinges throughout the stretching process, which requires the hinges to yield before buckling. However, the phenomenon can also occur more generally for hinges that plastify after buckling. 

\begin{figure*}[htb!]
    \centering
\includegraphics[width=0.98\linewidth]{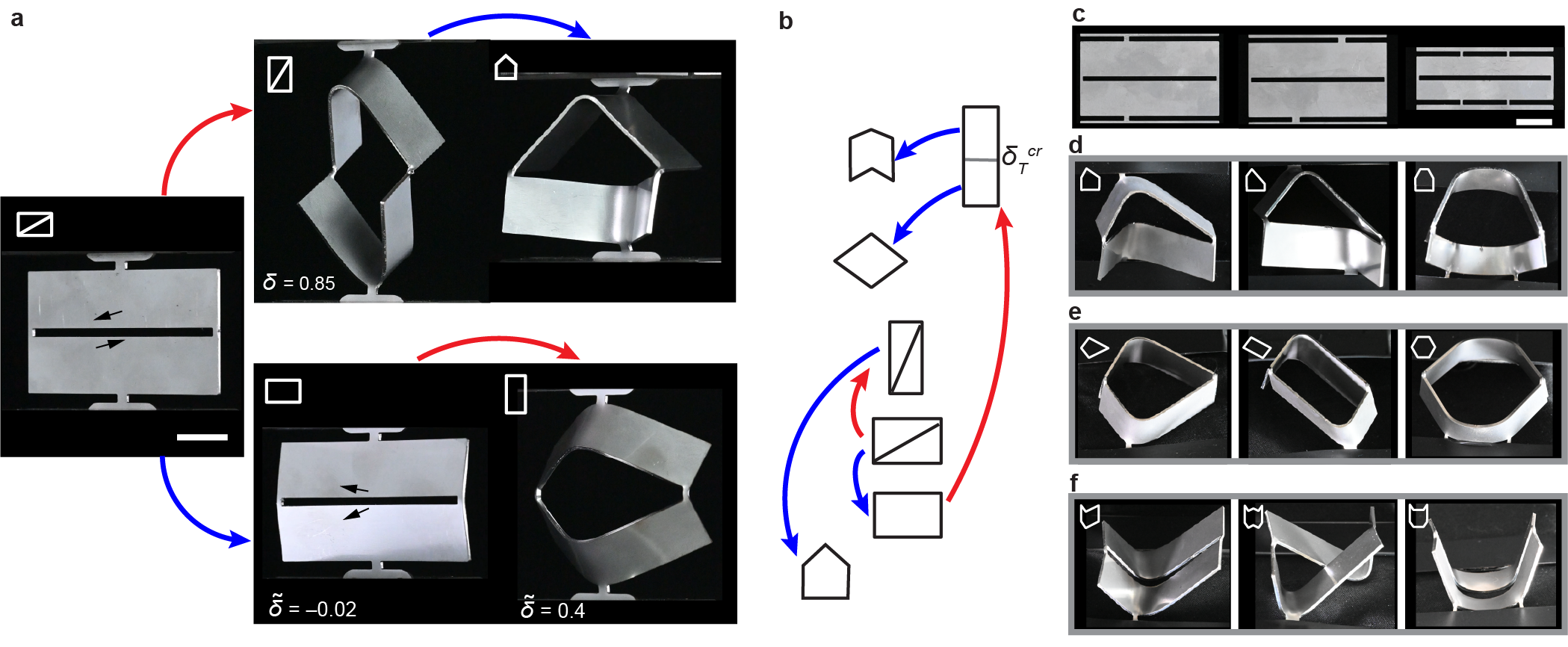}
\caption{\textbf{Additional self-guided folding shapes from a single kirigami.} \textbf{a,} Snapshots of kirigami with an initial antisymmetric imperfection (antisymmetric black arrows) under antisymmetric buckling and folding and under symmetric buckling. \textbf{b,} The transition graph indicates three morphed configurations from a single initial geometry under corresponding loading paths. \textbf{c-f,} Other types of kirigami structures and the corresponding folding shapes. \textbf{c,} Kirigami unit cells with two loading ligaments on one side (left), on the opposite side (middle), and with four loading ligaments (right). \textbf{d,} Pairs of corresponding isomorphisms of the configurations \protect\House, \protect\Square, and \protect\V, respectively. (d) isomorphism of \protect\House: \protect\HS, \protect\HA, and \protect\HT, (e) isomorphism of \protect\Square: \protect\SD, \protect\SR, and \protect\SH, and (f) isomorphism of \protect\V: \protect\VT, \protect\VA, and \protect\VU.  The scale bar is 15 mm.
}
    \label{fig:3 more morphing patterns}
\end{figure*}
The deformation angles given by \eqref{stretching angle} are frozen by plasticity during stretching. During the subsequent stages, these angles induce torques that guide folding. The kirigami folds into  \protect\Square\ when the angle $\theta$ increases and folds into \protect\V\ when $\theta$ decreases (Fig.~\ref{fig:1 folding shapes}b left). We apply force balance and torque balance at the rotating and bending joints under stretching and compression, and find the critical condition between \protect\Square\ and \protect\V\ as $\frac{d\theta}{d\delta_C}=0$ (see Supplementary Information), which leads to 
\begin{equation}
\tilde{\delta}_T^{cr} \; = \; 
\frac{3\,\tilde{b}\,\tilde{\ell}^3\left(1 + 2\tilde{b}\tilde{\ell}\right)}
     {\,2\tilde{\ell}^2 + \left(1 + \tilde{b}\tilde{\ell}\right)^3\,}.
     \label{foding boudanry}
\end{equation}
When $\frac{d\theta}{d\delta_C}>0$, or equivalently $\tilde\delta<\tilde{\delta}_T^{cr}$, the kirigami folds into \protect\Square. This inequality holds when the kirigami is stretched either under a small stroke or has a large hinge width $\tilde{b}=b/\ell_y$ and large aspect ratio $\tilde{\ell}=\ell_x/\ell_y$; otherwise, the kirigami folds into \protect\V. To verify this prediction, we perform finite-element simulations by varying the stretching stroke $\tilde{\delta}=\delta_T/2\ell_y$, the ligament width $\tilde{b}$, and the aspect ratio $\tilde{\ell}$ (Fig.~\ref{fig:1 folding shapes}c, see experiments with varying $\tilde{b}$ in Methods). The measurements confirm the prediction and indicate that the threshold of stretch between the \protect\Square\ and \protect\V\ can be tuned by the hinge width $\tilde{b}$ and aspect ratio $\tilde{\ell}$, and when $\tilde{b}$ and $\tilde{\ell}$ are large enough, only the \protect\Square\ mode is possible. We summarize the above two loading paths with a transition graph ~\cite{bense2021complex, liu2024controlled, melancon2022inflatable}, which includes the four states (Fig.~\ref{fig:1 folding shapes}d, \protect\SI, \protect\SB, \protect\Square, and \protect\V). This transition graph recapitulates how two distinct folded shapes can be obtained by a suitable stretching-compression protocol, guided by our theory. This theory is restricted to purely plastic hinges, which have yielded prior to buckling (See Methods). However, self-guided plastic folding will generally occur whenever buckling and plasticity are present, and is therefore broadly applicable for a wide range of kirigami geometries and elastoplastic materials.

 \begin{figure*}[t!]
    \centering
\includegraphics[width=1.\linewidth]{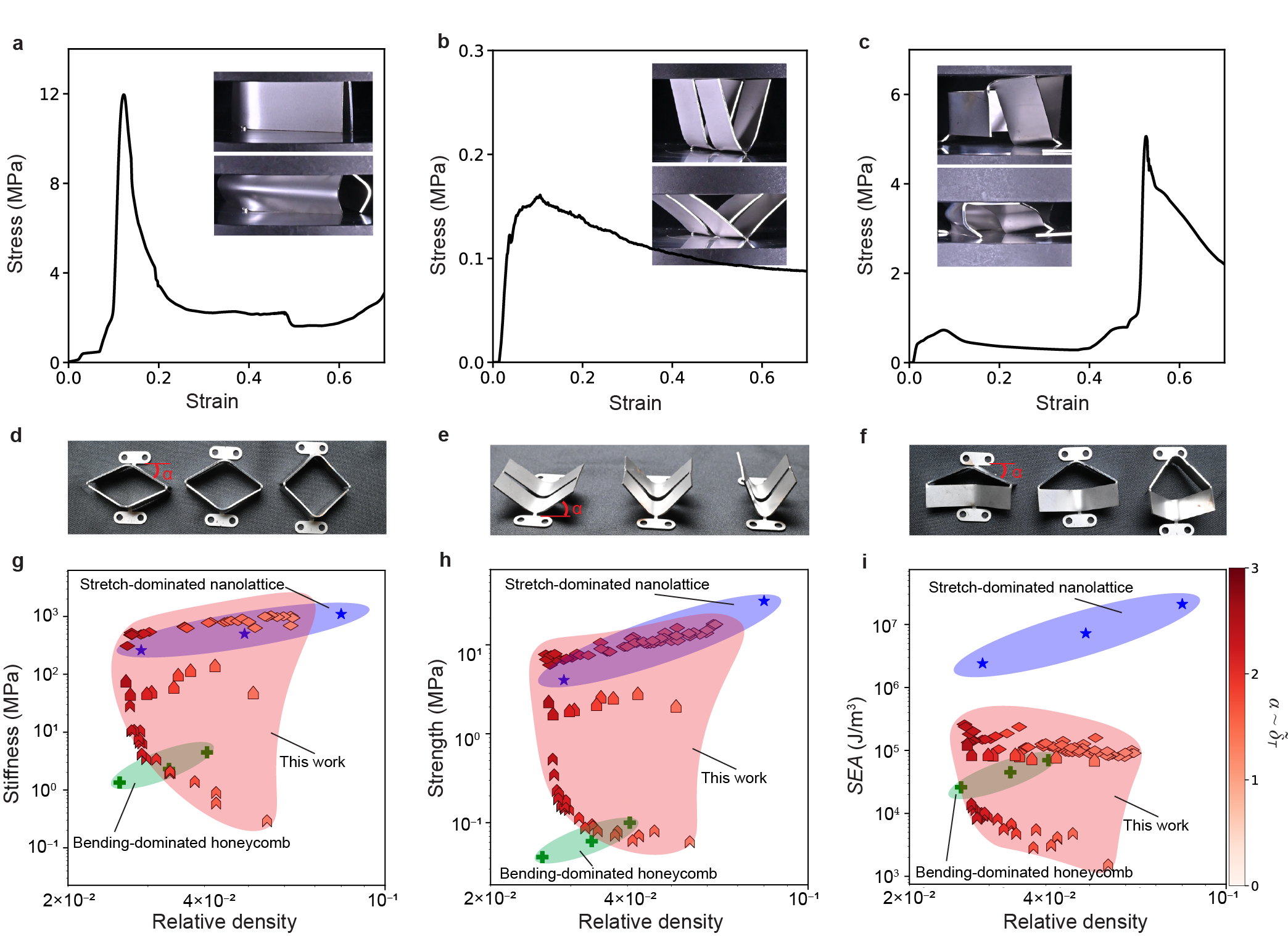}
\caption{\textbf{Mechanical performance under out-of-plane compression.} \textbf{a-c,} Stress-strain curves of the (a) \protect\Square, (b) \protect\V, and (c) \protect\House\ configurations and the corresponding snapshots (insets) under quasi-static compression. \textbf{d-f,} Snapshots of shapes \protect\Square, \protect\V, and \protect\House\ at different angle of $\alpha$ of different stretches. \textbf{g-i,} The compression stiffness (g), strength (h), and specific energy absorption (SEA) (i) versus relative density of the kirigami in this work (color density indicates the amount of stretch) compared with other existing load-bearing metamaterials of stretch-dominated nanolattices~\cite{meza2014strong} and bending-dominated honeycomb in-plane~\cite{papka1998experiments}.
}
    \label{fig:4 mechanical performance}
\end{figure*}

\section*{More shapes}
We have shown that a kirigami unit cell under symmetrical buckling can self-guide fold into \protect\Square\ and \protect\V\ shapes. We now show that it is possible to achieve more shapes by utilizing additional buckling modes readily present in the kirigami. The antisymmetric buckling mode of the kirigami is a higher buckling mode but can be triggered by embedding a transverse imperfection (see \protect\AI\ and \protect\AB\ in Fig.~\ref{fig:3 more morphing patterns}a and  Methods). After unloading back, the 3D antisymmetric shape folds itself towards a well-defined \protect\House\ shape (Fig.~\ref{fig:3 more morphing patterns}a). 
If one wishes to instead trigger the symmetrical buckling from the same antisymmetric initial state, we can simply compress it first. This triggers an additional compressive symmetric buckling mode (Fig.~\ref{fig:3 more morphing patterns}a). This mode, in fact, seeds a symmetric imperfection. In turn, this imperfection brings back the symmetric buckling of the kirigami once we stretch it (Fig.~\ref{fig:3 more morphing patterns}a and Supplementary Movie 2). As described earlier, such a symmetrically buckled shape can then fold into the \protect\Square\ or \protect\V\ depending on the value of stretch (Fig.~\ref{fig:1 folding shapes}a). 
In summary, a kirigami with a parallel pattern of cuts can adopt three different shapes depending on its loading history (see the transition graph in Fig.~\ref{fig:3 more morphing patterns}b). 

One can further enrich the range of shapes one can obtain. We do so
by varying the position and the number of the hinges (Fig.~\ref{fig:3 more morphing patterns}c-f and Supplementary Movie 2). As a result, we can obtain a range of kirigami structures that can morph into various shapes when stretched and compressed, such as \protect\HS, \protect\HA, and \protect\HT\ under antisymmetric buckling (Fig.~\ref{fig:3 more morphing patterns}d), \protect\SD, \protect\SR, and \protect\SH\ for small stretch of symmetrical buckling (Fig.~\ref{fig:3 more morphing patterns}j) and \protect\VT, \protect\VA, and \protect\VU\ for large stretch (Fig.~\ref{fig:3 more morphing patterns}k), respectively. This form of self-guided folding---where a single geometry can morph into multiple shapes in a simple loading condition---is distinct from earlier approaches---where one typically obtains a single shape~\cite{rafsanjani2017buckling} or multiple shapes resorting to manual control \cite{overvelde2017rational, silverberg2014using} or complex automated control~\cite{chen2021reprogrammable}.

\section*{Tunable rigidity}
Conventional load-bearing structures~\cite{meza2014strong, liu2024harnessing} exhibit a single load-bearing capacity defined by a fixed initial geometry. Multistable metamaterials address this challenge somewhat, yet they exhibit a finite number of stable configurations. In addition, these multistable materials~\cite{liu2024controlled,kamp2025reprogrammable, meng2022deployable, fu2018morphable} are often made from soft materials and morph into compliant geometries. In contrast, our metamaterials are continuously tunable within a single configuration and enable multiple distinct configurations from stretching-dominated \protect\Square\ to bending-dominated \protect\V\ under out-of-plane compression. Therefore, they exhibit tunable load-bearing capacity.

To demonstrate their high performance, we conduct out-of-plane compression on the three folding configurations (\protect\Square, \protect\V, and \protect\House) with different stretches (Fig.~\ref{fig:4 mechanical performance}a-c and Supplementary Movie 3). The \protect\Square\ configuration exhibits the highest stiffness and strength at initial compression due to a stretch-dominated mechanism when the plates are loaded along their axes. As the compression increases, the two half-\protect\Square\ plates start to buckle and fold, followed by a significant force drop (Fig.~\ref{fig:4 mechanical performance}a). The deformation of the \protect\V\ configuration is bending-dominated, resulting in the lowest stiffness and strength and a stable stress plateau during compression (Fig.~\ref{fig:4 mechanical performance}b). The \protect\House\ configuration combines the features of the \protect\Square\ and \protect\V\ configurations. The half-\protect\V\ component of the \protect\House\ shape deforms first, showing a low-force plateau in the first stage of compression. In the second stage of compression, the structure exhibits a much higher strength and stiffness after the half \protect\Square\ component enters into contact (Fig.~\ref{fig:4 mechanical performance}c). Hence, the three distinct shapes exhibit markedly different deformation modes, which endow them with a vast range of mechanical performances.

\begin{figure*}[htb!]
    \centering
\includegraphics[width=0.98\linewidth]{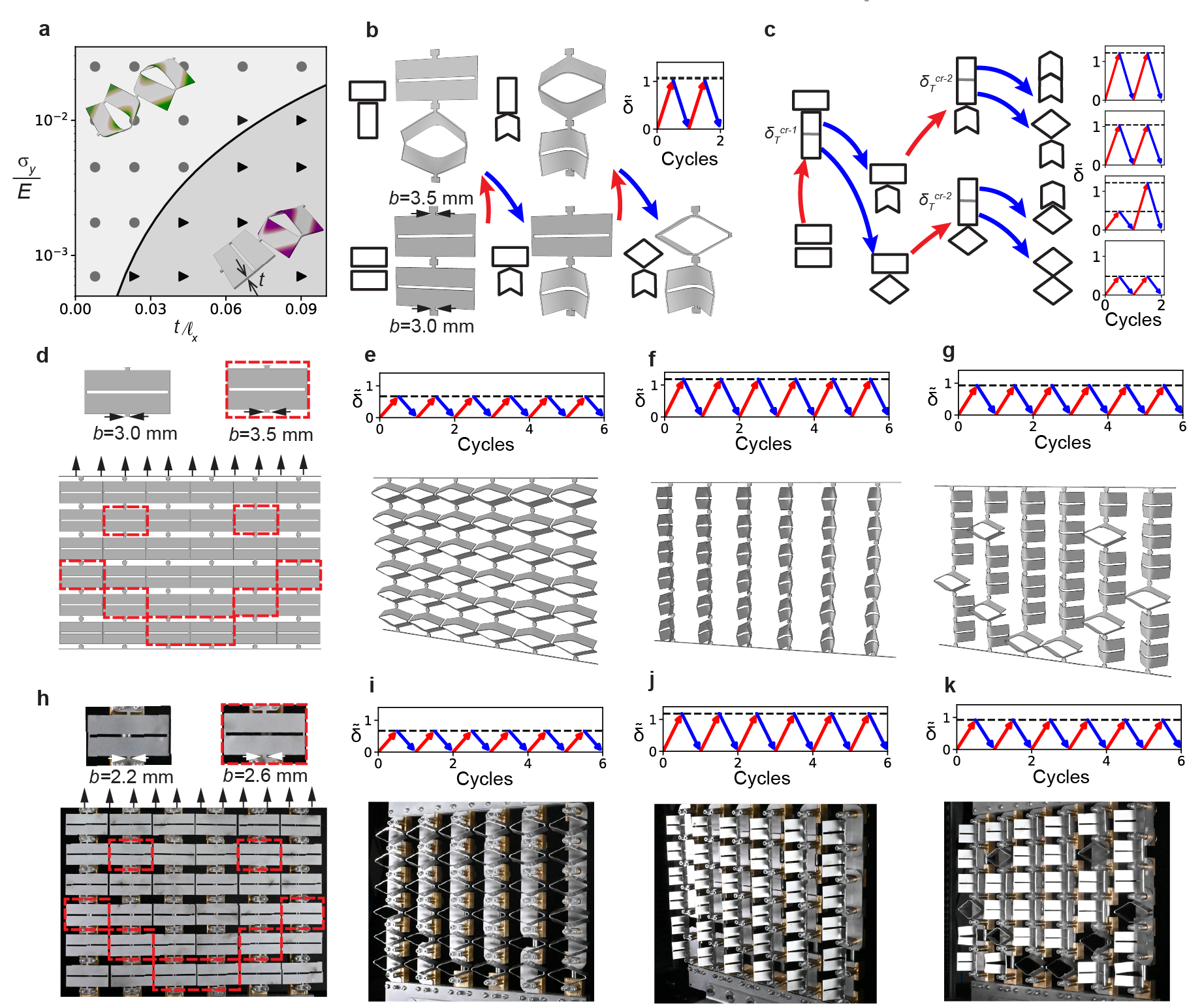}
\caption{\textbf{Shape morphing kirigami programmed by loading history.} \textbf{a,} Mode of buckling vs. plastic strain $\sigma_y/E$ and ratio of $t/\ell_x$. The solid line denotes the theoretical boundary between elastic buckling and yield buckling (Eq.~\eqref{eq:yield buckling}). The triangles and circles are simulation results that denote localized and simultaneous buckling, respectively, in a two-unit kirigami structure. The bottom-right inset has values of $\sigma_y/E=0.00175$ and $t/\ell_x=0.0431$. The top-left inset has values of $\sigma_y/E=0.0045$ and $t/\ell_x=0.0235$. \textbf{b,} Snapshots of sequential self-guided folding of a two-unit kirigami under two cycles of loading history, where we make the bottom unit buckle first. \textbf{c,} Transition graph of four morphing shapes (\protect\VV, \protect\SV, \protect\VS, and \protect\SSS) from the two-unit kirigami under four different loading histories. \textbf{d-k,} Multiple morphing patterns with sequential self-guided folding in a kirigami programmed by loading histories where (d-g) are FEM results and (h-k) are experiments. \textbf{d} and \textbf{h,} A $6\times6$ kirigami at rest consists of two types of unit cells. \textbf{e} and \textbf{i,} The first morphing pattern under six loading cycles of  $\tilde{\delta}=0.67$. \textbf{f} and \textbf{j,} The second morphing pattern under six loading cycles of $\tilde{\delta}=1.2$. \textbf{g} and \textbf{k,} The third morphing pattern under six loading cycles of $\tilde{\delta}=0.93$. The red (blue) arrows indicate stretching (unloading back) (see Supplementary Movie 4).
}
    \label{fig:2 kirigami metamaterials}
\end{figure*}
We then quantify such performances of the three configurations at varying stretch levels (described by angle $\alpha$, see $\alpha$ increases as stretch in Fig.~\ref{fig:4 mechanical performance}d-f) by plotting the Ashby map of compressive strength, stiffness, and specific energy absorption (SEA) versus the relative density (see definition in Methods). 
The \protect\Square\ and the \protect\V\ configurations stay in the top and bottom regimes for the entire range of relative density (stretches), indicating the highest and the lowest values, respectively, of stiffness and strength (Fig.~\ref{fig:4 mechanical performance}g and h). The \protect\House\ configuration exhibits values that lie between those of the \protect\V\ and \protect\Square\ configurations. Interestingly, while the stiffness and strength of the \protect\House\ configuration are an order of magnitude lower than those of the \protect\Square\ configuration, its specific energy absorption is the same order to that of the \protect\Square\ configuration (Fig.~\ref{fig:4 mechanical performance}i). The enhanced energy absorption of the \protect\House\ configuration arises from the coupling between the half-\protect\Square\ and half-\protect\V\ components under compression. As the relative density decreases (stretch increases), stiffness, strength, and energy absorption highly increase in the \protect\V\ configuration. The stretch value has less effect on the properties of the \protect\Square\ and the \protect\House\ configurations. We find that, across these performance metrics, our kirigami, from the \protect\V\ to the \protect\Square\ configuration, exhibits a wide range of load-bearing capacity spanning two to three orders of magnitude, from the compliant, bending-dominated honeycomb lattices~\cite{papka1998experiments} to the extremely stiff and strong nanoarchitectures~\cite{meza2014strong}. Existing stretch-dominated nanoarchitectures and bending-dominated honeycomb lattices exhibit distinct mechanical performances that arise from their complex initial geometries. In contrast, our kirigami with varying stretches exhibits comparable mechanical performance from a single 2D geometry with plasticity, thus could be readily produced at a wide range of scales with a wide range of materials---from any elastoplastic sheet.

\section*{From the unit cell to the metamaterial}
So far, our study has focused on kirigami unit cells. We conclude the paper by demonstrating that our approach can be generalized to kirigami metamaterials, where the number of available shapes explodes.
%We have identified the stretching thresholds between the two symmetrical configurations for a given initial kirigami geometry. 
%We next leverage them (Fig.~\ref{fig:1 folding shapes}c and d) to design kirigami metamaterials that exhibit on-demand morphing patterns that can be chosen by the loading protocol. 
To do this, we first engineer buckling sequences---layer by layer---in our kirigami. To this end, we generalize the concept of yield buckling~\cite{liu2024harnessing, liu2025tuning, liu2026combinatorial} to kirigami. Unlike elastic buckling in a kirigami strip, where all the unit cells buckle simultaneously, yield buckling leads to localization due to negative stiffness (softening) at the onset of buckling (see inserts of Fig.~\ref{fig:2 kirigami metamaterials}a and Methods). Yield buckling occurs when the following inequality is satisfied:
%which enables a buckling in sequence, 
\begin{equation}\label{eq:yield buckling}
    \frac{\sigma_y}{E}\leq
    \left[\frac{\ell_y^2}{2b\ell_x}+\frac{1}{4(1-\nu)}\right]\left(\frac{t}{\ell_x}\right)^2.
\end{equation}
$\sigma_y$ is the yield stress, $E$ is the Yong's modulus, $\nu$ is the Poisson's ratio, and $t$ is the kirigami thickness. The left-hand side of Eq.~\eqref{eq:yield buckling} is the yield load of the hinges, and the right-hand side represents the elastic buckling load (see Methods). %Unlike elastic buckling in a kirigami strip, where all the unit cells buckle simultaneously, yield buckling leads to localization due to negative stiffness (softening) at the onset of buckling (see inserts of Fig.~\ref{fig:2 kirigami metamaterials}a and Methods). 
In postbuckling, the load continues to decrease until the hinges no longer bend and start to stretch. At this point, the kirigami notably stiffens (see Methods). The combination of yield buckling and stiffening mechanisms leads to sequential buckling in a kirigami strip under continuous stretching (see Methods and Supplementary Movie 4) and enables sequential self-guided folding under stretch-compression loading cycles (see Methods and Supplementary Movie 4). 
We then demonstrate this idea in a kirigami consisting of two lumped unit cells with different stretch thresholds (Fig.~\ref{fig:2 kirigami metamaterials}b). We make the bottom unit buckle first under initial stretching. It folds into the \protect\V\ under compression. To avoid global out-of-plane deflection during compression, we only allow the movement of the middle joints along the uniaxial direction (see the sliding blocks along a column in the experiments in Methods). We then freeze the \protect\V\ shape by applying a fixed constraint between the joints of the unit cell (see a hook structure locked between sliding blocks in experiments in Methods), which stiffens the structure and enables the top unit to buckle and fold into the \protect\Square\ in the second loading cycle and the kirigami to fold into \protect\SV\ (Fig.~\ref{fig:2 kirigami metamaterials}b). Such a two-cell kirigami enables four different morphing combinations (see \protect\SSS, \protect\SV, \protect\VS, and \protect\VV\ in Fig.~\ref{fig:2 kirigami metamaterials}c), where three of them (\protect\SSS, \protect\SV, and \protect\VV) have a periodic loading history and are independent to the buckling order. For a 2D kirigami metamaterial with periodic loading history, the maximum number of morph patterns is $m+1$, where m is the types of unit cells of different stretch thresholds ($m=2$ in Fig.~\ref{fig:2 kirigami metamaterials}b and c).
In the next, we will focus on the periodic loading history and demonstrate multiple programmable morphing patterns in a single 2D kirigami.   

We numerically and experimentally demonstrate this idea in a $6\times6$ kirigami consisting of two types of unit cells, where each column of units is independent of the others (see Fig.~\ref{fig:2 kirigami metamaterials}d-h for numerics and Fig.~\ref{fig:2 kirigami metamaterials}h-k for experiments). We fix the bottom side of the kirigami and apply six loading cycles to the other side. For the first morphing pattern, we apply a relatively small stretch ($\tilde{\delta}=0.67$) in each cycle (Fig.~\ref{fig:2 kirigami metamaterials}e top and i top) where all unit cells fold into \protect\Square\ 
%, transforming the kirigami into a \protect\Square\ lattice 
 (Fig.~\ref{fig:2 kirigami metamaterials}e bottom and i bottom and Supplementary Movie 4). In the second case, we significantly increase the stretch stroke of each cycle ($\tilde{\delta}=1.2$), where all unit cells fold into \protect\V\ 
%, transforming the kirigami into a \protect\V\ lattice 
 (Fig.~\ref{fig:2 kirigami metamaterials}f and j and Supplementary Movie 4). Finally, we apply a stretch stroke between the two cases described above ($\tilde{\delta}=0.93$). The unit cells in the red boxes fold into the \protect\Square\ configuration while the rest fold into the \protect\V\ configuration, which morphs the metamaterial to a smiley pattern (Fig.~\ref{fig:2 kirigami metamaterials}g and k and Supplementary Movie 4). 
As we increase the number of unit cells $n$ with different stretch thresholds in the design of metamaterials, we can program $n+1$ morphing patterns within a single metamaterial using periodic loading cycles (see Methods for four morphing patterns from a single kirigami). 

% In conclusion, self-guided plastic folding enables shape-morphing of a large number of 3D shapes ranging from bending-dominated to stretching-dominated from the same 2D sheet. Those 3D shapes then exhibit widely tunable mechanical properties. 

In conclusion, self-guided plastic folding enables a single 2D sheet to transform into diverse 3D architectures, ranging from bending-dominated to stretching-dominated, with widely tunable mechanical properties. This strategy is applicable across different length scales and elastoplastic materials.
Self-guided plastic folding could therefore be translated to a range of applications, from architecture, aerospace, automotive, and robotics at the meter scale to wearable electronics and medical implants at the millimeter scale, and to reconfigurable lenses at the micrometer scale. Beyond these applications, we anticipate that self-guided plastic folding could provide a versatile platform for exploring programmable shape transformations and adaptive mechanical functionality in soft robotics, mechanical metamaterials, and photonic metasurfaces. Looking ahead, we envision the full exploration of the process-design map, generalization to more complex shapes, free-constraint kirigami benefits from three-dimensional design, and resettable morphing patterns to be formidable future research directions.\\

% \subsection*{Supporting Information Appendix (SI)}
% \textbf{Data, Materials, and Software Availability.} 
All the codes and data supporting this study are available on the public repository https://zenodo.org/records/24021744\\\\

\noindent\textbf {Acknowledgment}: We thank M. van Hecke and J.T. Overvelde for insightful discussions and suggestions, and C. Ederveen Janssen and E. Hop for technical assistance. We acknowledge financing from the European Research Council under grant agreements 852587, 101212785 and 101170693, and the Netherlands Organization for Scientific Research under grant agreement NWO TTW 17883.
\bibliographystyle{unsrt}
\bibliography{refs.bib}
\clearpage
\onecolumngrid
\subsection*{Appendix: Methods}

We first exploit the elastic buckling modes of the kirigami unit cell and the regime of yield buckling in the presence of plasticity. We then introduce sequential yield buckling and sequential self-guided folding in kirigami metamaterials. Finally, we present the scalability of self-guided folding, experimental protocols, and finite-element simulation protocols.

\subsection*{Symmetric and antisymmetric buckling}
\begin{figure*}[htb!]
\centering\includegraphics[width=0.9\linewidth]{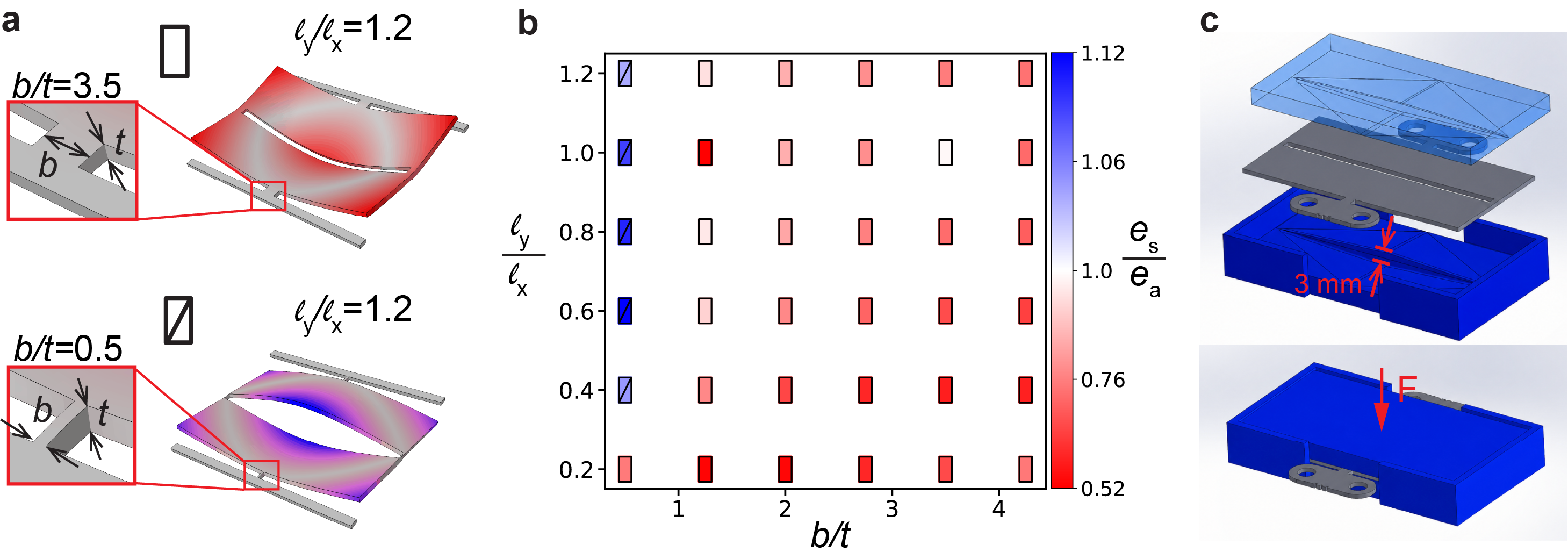}
\caption{\textbf{Symmetric and antisymmetric buckling.} 
\textbf{a,} Snapshots of the first buckling mode of unit cells at ratios of $b/t=3.5$ and $\ell_y/\ell_x=1.2$ (symmetric, top) and the ratios of $b/t=0.5$ and $\ell_y/\ell_x=1.2$ (antisymmetric, bottom).
\textbf{b,} Eigenvalue ratio $e_s/e_a$ between symmetric and antisymmetric buckling vs. the unit-cell length ratio $\ell_y/\ell_x$ and the ligament ratio $b/t$. \protect\SB\ ($e_s/e_a<0$) and \protect\AB\ ($e_s/e_a>0$) denote that the first buckling mode is symmetric and antisymmetric, respectively. \textbf{c,} Technique used to seed an antisymmetric imperfection. A flat kirigami pressed into a pair of molds with an antisymmetric surface.}
    \label{fig:S2}
\end{figure*}

Two possible buckling modes can be observed in a kirigami with parallel cuts under stretching: symmetric and antisymmetric buckling ( Fig.~\ref{fig:S2}a). By analyzing the elastic buckling load under symmetric and antisymmetric modes, we find that the antisymmetric buckling load $\tilde{F}^a_{cr}$ is always larger than the symmetric buckling load $\tilde{F}^s_{cr}$. 
\begin{equation}
\tilde{F}^a_{cr}\approx\tilde{F}^s_{cr}+\frac{4J}{ht^3}
\end{equation}
where $J$ is the second polar moment of the hinge section.
Only in the case where $b \ll 2t$,  $J \approx \frac{1}{24} tb^3$, and $\tilde{F}^a_{cr} \approx \tilde{F}^s_{cr}$, can the antisymmetric mode be possible as the first buckling mode.

 To verify this prediction, we conducted numerical simulations with varying geometric ratios $b/t$ and $\ell_y/\ell_x$. By measuring the eigenvalue ratio $e_s/e_a$ between the symmetric and antisymmetric modes (Fig.~\ref{fig:S2}b) of each geometry, we find that the symmetric buckling mode is dominant. Although the antisymmetric mode is possible for small $b/t$ in the linear-elastic regime, kirigami hinges usually fracture before buckling occurs in practice. To trigger the antisymmetric buckling for an elastoplastic kirigami, we can compress the kirigami unit cell out of plane in a model with an antisymmetric texture (Fig.~\ref{fig:S2}c and Fig.~\ref{fig:3 more morphing patterns}a). Next, we will focus on the self-guided folding behavior in the presence of plasticity based on symmetric buckling.

\subsection*{Yield buckling and sequential buckling in kirigami}
\begin{figure*}[htb!]
    \centering
\includegraphics[width=0.95\linewidth]{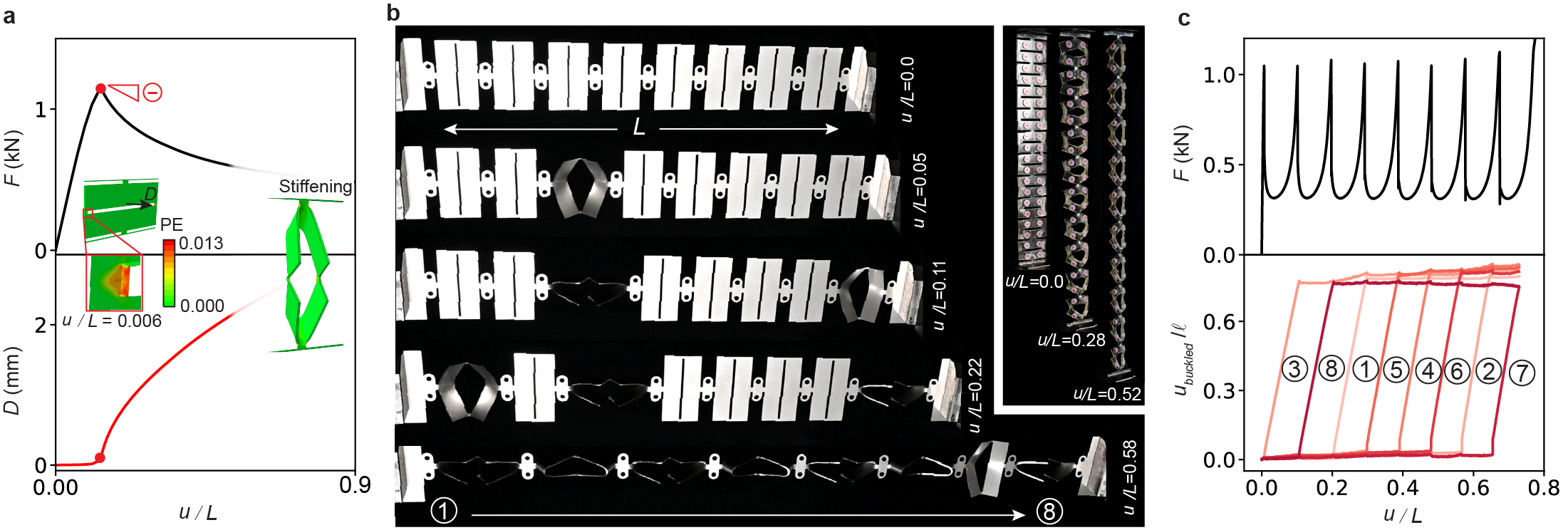}
\caption{\textbf{Sequential yield buckling in kirigami.}  \textbf{a,} Numerical force $F$ and deflection $D$ versus stretching stroke $u/L$ curves of a kirigami unit cell. The insets show the plastic strain (PE) state of the unit cell at the moment of buckling and the unit cell configuration at the stiffening state. \textbf{b,} Snapshots of a kirigami strip with eight unit cells laser-cut from a 316L stainless steel plate under stretching. The top-right inset shows snapshots of a 3D-printed elastic kirigami strip under varying stretches. \textbf{c,} Force $F$ and buckling stroke of each unit cell $u_{buckled}/\ell$ versus stretching stroke $u/L$ curves of the steel kirigami. The number of unit labels indicates the buckling order of each unit cell.}
    \label{fig:S3}
\end{figure*}

Buckling-induced kirigami has been widely studied in elastic~\cite{yang2018multistable, janbaz2024diffusive} and in elastoplastic sheets~\cite{rafsanjani2017buckling}, where kirigami typically buckles before the hinges plastify, i.e., elastic buckling. Such kirigami with elastic buckling exhibits positive stiffness at the onset of buckling and triggers all periodic unit cells to buckle simultaneously. Here, we consider buckling in the presence of plasticity and the collective buckling behavior in a kirigami strip. We introduce the concept of `yield buckling' into the kirigami unit cell~\cite{liu2024harnessing}, in which hinges yield before buckling occurs, resulting in negative stiffness at the onset of buckling (Fig.~\ref{fig:S3}a). 
Unlike elastic buckling, yield buckling with negative stiffness naturally leads to localization~\cite{liu2024harnessing}. From the theory of elastic buckling and plastic yielding, we can define the regime of yield buckling as 
\begin{equation}
    \frac{\sigma_y}{E}\leq
    \left[\frac{\ell_y^2}{2b\ell_x}+\frac{1}{4(1-\nu)}\right]\left(\frac{t}{\ell_x}\right)^2.
\end{equation} 
which is Equation~\eqref{eq:yield buckling} of the Main Text.
To verify this prediction, we conduct numerical simulations by varying the parameters $\sigma_y/E$ and $t/\ell_x$ in a two-cell kirigami (Fig.~\ref{fig:2 kirigami metamaterials}a in the Main Text). The results confirm that only yield buckling can separate the buckling modes of unit cells in series.  

The load drop at the onset of buckling will continue deep into the postbuckling regime when the kirigami is further stretched until the deformation of hinges changes from bending to stretching, where the kirigami notably stiffens (Fig.~\ref{fig:S3}a, right insert). The conjunction of yield buckling and the stiffening mechanism is the key to achieving sequential buckling when the unit cells are connected in series~\cite{liu2024harnessing}.

We then experimentally demonstrate sequential buckling in a strip of kirigami with eight periodic unit cells of $\sigma_y/E=0.0015$ and $t/\ell=0.039$ (Fig.~\ref{fig:S3}b and Supplementary Movie 4). Under quasi-static stretching, the kirigami first exhibits high stiffness before buckling occurs. As the stretch increases, the third unit cell starts to buckle at a stroke of 0.008, followed by a negative stiffness and a significant force drop at the onset of buckling (Fig.~\ref{fig:S3}b and c top). The unit cell stiffens when the hinges and plates of the unit cell start to stretch, and when the load exceeds the yield buckling load, the next step of buckling is triggered (Fig.~\ref{fig:S3}b). The same buckling behaviors repeat for the remaining unit cells, contributing eight steps of sequential buckling until a stroke of 0.8 with a random buckling order (Fig.~\ref{fig:S3}b). By contrast, a kirigami strip made of an elastic material undergoes simultaneous deformation in the stretch (Fig.~\ref{fig:S3}b inset). Thus, sequential yield buckling in kirigami combines stiffness and energy dissipation, and therefore enables the design of effective shock absorbers under stretching loading conditions (Fig.~\ref{fig:S3}c). In addition, such sequential yield buckling enables the sequential self-guided folding of the \protect\Square\ and the \protect\V.

\subsection*{Sequential self-guided folding}
\begin{figure*}[htb!]
    \centering
\includegraphics[width=0.99\linewidth]{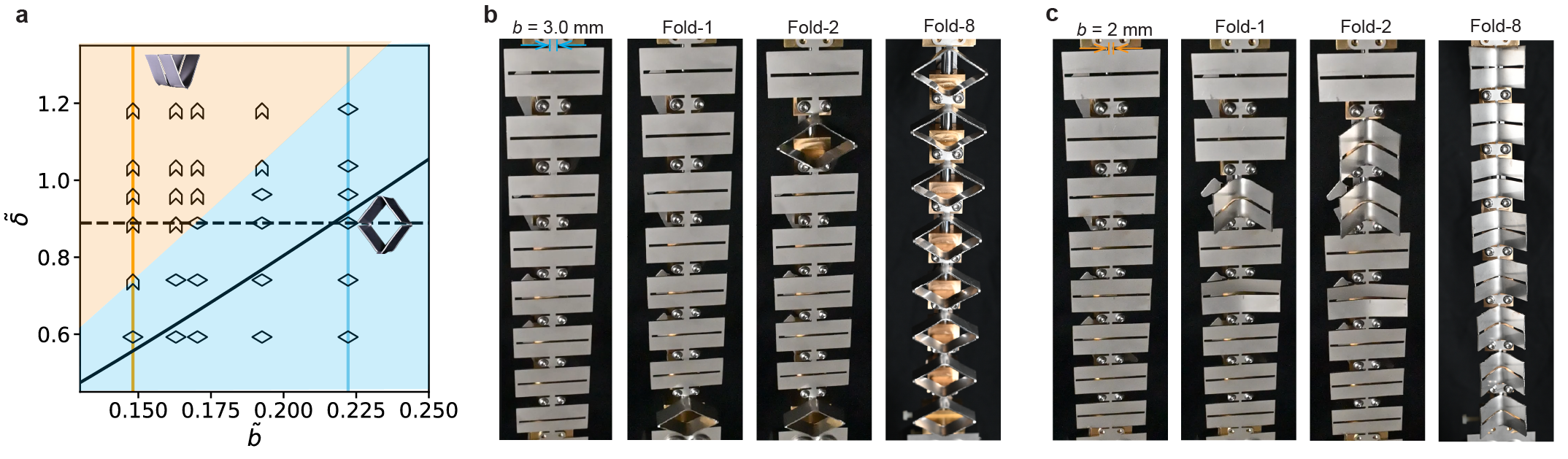}
\caption{\textbf{Sequential self-folding kirigami under eight cycles of stretching and compression.} \textbf{a,}  Stretch $\tilde{\delta}=\delta_T/2\ell_y$ vs. normalized hinge size $\tilde{b}=b/\ell_y$. The markers \protect\Square\ and \protect\V\ are experiments. The black solid line is Eq.~\eqref{foding boudanry} for a given $\tilde{\ell}=1.86$, respectively. \textbf{b and c,} Snapshots of a strip of kirigami consisting of eight unit cells with ligament ratio $b=3.0$ (b) and $b=2.0$ (c).
}
    \label{fig:S4}
\end{figure*}

We have shown that a kirigami strip of eight unit cells can buckle sequentially under stretching. Here, we experimentally demonstrate that such a kirigami can sequentially buckle and fold into a strip of \protect\Square\ and \protect\V\ lattices under cyclic stretching and compression. To prevent global deflection under compression, we connect joints between adjacent unit cells on the sliding blocks with only movement in the stretching direction (see Fig.~\ref{fig:setup} for details). Before proceeding, we first experimentally measure the critical stretching thresholds of the kirigami unit cell as a function of hinge width $\tilde{b}$ (markers in Fig.~\ref{fig:S4}a). We find that the experimentally determined thresholds are higher than the predictions of linearized theory (black solid line in Fig.~\ref{fig:S4}a). This discrepancy may arise from material nonlinearities at large strains in the experiments.

In the first strip of kirigami, we use wide hinges of $\tilde{b}=0.21$ ($b = 3$ mm), which enable a robust \protect\Square\ configuration over a wide range of stretch (Fig.~\ref{fig:S4}a sky-blue line).
In the first cycle of stretch and compression, the bottom unit cell first buckles and folds to a \protect\Square. The \protect\Square\ shape is automatically locked by a hook structure at the back of the slider, which notably stiffens the kirigami strip and enables the second buckling and folding process (Fig.~\ref{fig:S4}b). The rest of the unit cells repeat the same buckling-folding process and sequentially fold to a strip of \protect\Square\ (Fig.~\ref{fig:S4}b and Supplementary Movie 4). The same process can be used to achieve a strip of \protect\V\  lattice by choosing a small hinge of $\tilde{b}=0.14$ ($b =2$ mm) (Fig.~\ref{fig:S4}ac). 
We then generalize this strategy into 2D kirigami. 

\subsection*{Programmable shape-morphing metamaterials}
\begin{figure*}[htb!]
    \centering
\includegraphics[width=0.98\linewidth]{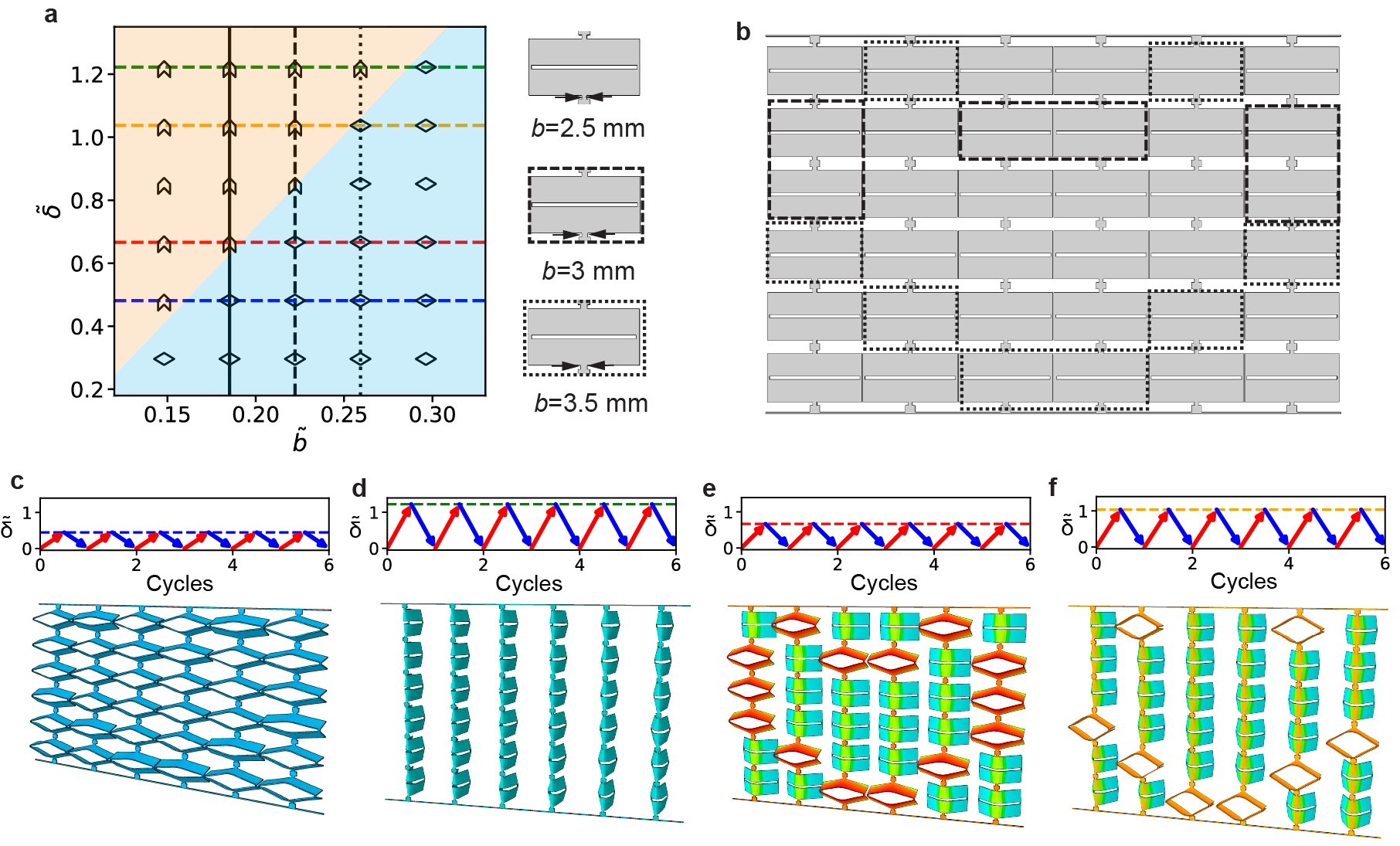}
\caption{\textbf{Numerical simulations of morphing patterns from heart to smiley.} \textbf{a,} Folding configurations of three types of unit cells under four different stretches. \textbf{b,} A kirigami of $6\times6$ unit cells of three different hinge widths. \textbf{c-f,} Four programmed morphing patterns of the kirigami folded from the four loading histories. (c) \protect\Square\ lattice under a load history of $\tilde{\delta}=0.48$. (d) `Heart' pattern under a load history of $\tilde{\delta}=0.67$. (e) `Smiley' pattern under a load history of $\tilde{\delta}=1.0$. (f) \protect\V\ lattice under a load history of $\tilde{\delta}=1.2$. 
}
    \label{fig:S5}
\end{figure*}
In the Main Text, we demonstrate that a 2D kirigami with $6\times6$ unit cells with two types of unit cells of different stretching thresholds can morph into three distinct patterns programmed by three different loading protocols. Here, we further numerically demonstrate that four distinct morphing patterns can be programmed in a single kirigami of three unit-cell types with different stretching thresholds (Fig.~\ref{fig:S5}ab). In the loading histories of the lowest and the highest stretch strokes (Fig.~\ref{fig:S5}c top and bottom), the kirigami sequentially buckles and folds into a \protect\Square\ lattice and \protect\V\ lattice, respectively (Fig.~\ref{fig:S5}de). For a loading history of stroke $\tilde{\delta}=0.67$ (Fig.~\ref{fig:S5}c), the morphing pattern switches to a `heart' (Fig.~\ref{fig:S5}f). The fourth `smiley' pattern appears when the stroke of loading history increases to $\tilde{\delta}=1.0$ (Fig.~\ref{fig:S5}cg).

\subsection*{Fabrication and scalability}

\begin{figure*}[htb!]
    \centering
\includegraphics[width=0.9\linewidth]{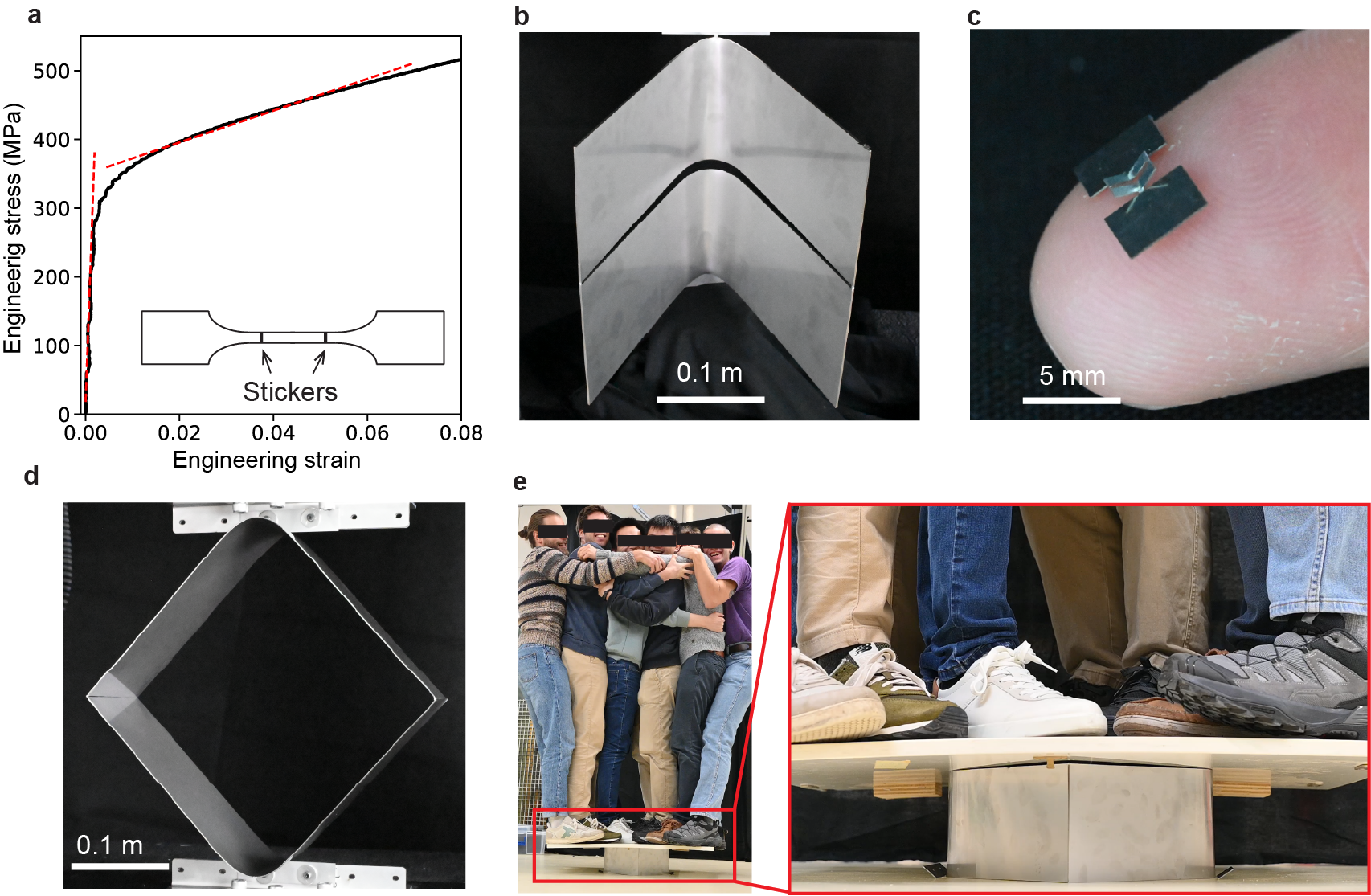}
\caption{\textbf{Material calibration and multi-scale production.} \textbf{a,} material calibration of 316L stainless steel. \textbf{b,} Snapshot of \protect\V\ configuration morphed from kirigami of size $456\text{ mm}\times267\text{ mm}\times 2\text{ mm}$ (width $\times$ length $\times$ thickness) under a stretch of 240\text{ mm}.  \textbf{c,} Snapshot of \protect\V\ configuration morphed from micro-kirigami of size $2.85\text{ mm}\times1.65\text{ mm}\times 0.1\text{ mm}$ under a stretch of 1.2\text{ mm}. \textbf{d,} Snapshot of \protect\Square\ configuration morphed from kirigami of size $456\text{ mm}\times267\text{ mm}\times 1\text{ mm}$ under a stretch of 240\text{ mm}. \textbf{e,} Load-bearing test of the \protect\Square\ shape in panel d under a weight of six adults. 
}
    \label{fig:S1}
\end{figure*}

The self-guided folding kirigami in this work relies on plasticity from a single material, thus is easy to produce at any scale and from readily available materials. Here, we use 316L stainless steel laser-cut from a plate of thickness $t$, which has Young's modulus $E=200$ GPa, yield stress $\sigma_y=300$ MPa, and tangent modulus $E_t=2$ GPa (see calibration in Fig.~\ref{fig:S1}a).  

We first prove the scalability of the mechanical self-guided folding in a half-meter-scale kirigami and in a micro-kirigami with the morphed configuration of the \protect\V\ (Fig.~\ref{fig:S1}b and c and Supplementary Movie 3), which may be of interest for the construction of load-bearing structures in multiple scales in a fast and robust way. We then demonstrate the load-bearing capacity of a half-meter \protect\Square\ shape that supports the weight of six adults (around 400 kg) on it (Fig.~\ref{fig:S1}d and e and Supplementary Movie 3). 

\subsection*{Kirigami unit cell undergoes multiple loading cycles}
\begin{figure*}[htb!]
    \centering
\includegraphics[width=0.9\linewidth]{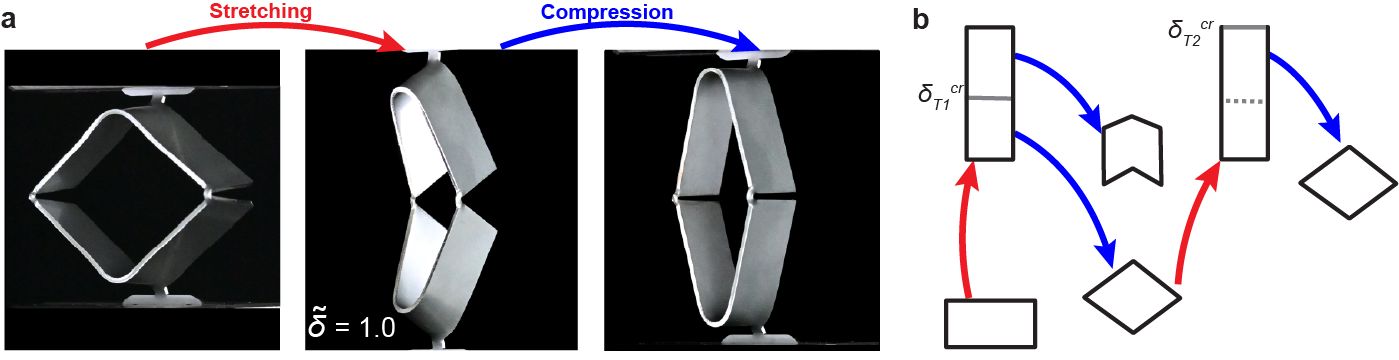}
\caption{\textbf{kirigami folding memory.} \textbf{a,} The \protect\Square\ in Fig.~\ref{fig:1 folding shapes}a of the Main Text undergoes a second loading cycle with a larger stretch. \textbf{b,} A transition graph of the kirigami contains four states of configurations and two cycles of loading paths. 
}
    \label{fig: second folding}
\end{figure*}

We have shown that a kirigami unit cell under a symmetric buckling mode can self-assemble into a \protect\Square\ in a small stretch and into a \protect\V\ in a larger stretch (Fig.~\ref{fig:1 folding shapes}a in the Main Text). Will the \protect\Square\ switch into the \protect\V\ under a second loading cycle of the larger stretch? To answer this question, we conduct a second loading cycle to the \protect\Square\ of a larger stretch $\tilde{\delta}=1.0$ (Fig.~\ref{fig: second folding}a). Interestingly, the unit cell under such a second stretch stays in the \protect\Square\ configuration under the second folding, which means the final folding configuration depends not only on the initial state of kirigami and the stroke of stretching, but also on the folding history. Such a folding memory arises from large plastic deformation at the hinges during the first folding. We use a transition graph to describe all the possible folding configurations under two loading cycles (in Fig.~\ref{fig: second folding}b). 

\subsection*{Geometry of folding kirigami}
\begin{figure*}[htb!]
    \centering
\includegraphics[width=0.9\linewidth]{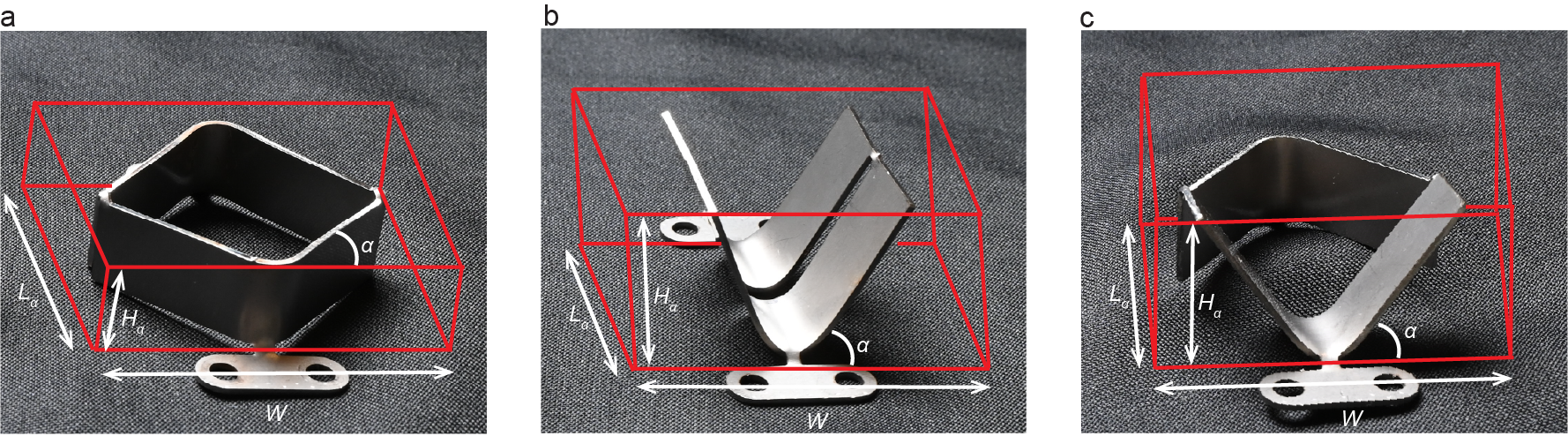}
\caption{\textbf{Relative density of folding kirigami morphed from a flat kirigami unit of width $W$, length $L$, and thickness $t$.} \textbf{a,} \protect\Square. \textbf{b,} \protect\V. \textbf{c,} \protect\House. The width of the 3D unit cell is the same as the initial 2D unit cell. Only the length $L_\alpha$ and height $H_\alpha$ change in the 3D unit cell.}
    \label{fig: relative density}
\end{figure*}

We have shown that a single kirigami can self-assemble into a stretch-dominated \protect\Square, a bending-dominated \protect\V, and a combined \protect\House, exhibiting significant tunability in load bearing. Here, we define the relative density of the three types of morphing shapes(Fig.~\ref{fig: relative density}) based on their corresponding lattice structures (Fig.~\ref{fig:2 kirigami metamaterials}ef). We consider a 2D kirigami consisting of a kirigami unit cell of width $W$, length $L$, and thickness $t$. For a \protect\Square\ lattice morphed from the 2D kirigami, the relative density can be defined as: $\Bar{\rho}\approx W\times L\times t/(L_\alpha\times W\times H_\alpha)\approx t/(W\sin\alpha)$. The relative density of a \protect\V\ lattice can be defined as: $\Bar{\rho}\approx 2t/(W\sin\alpha)$.  The relative density of a \protect\House\ lattice is: $\Bar{\rho}\approx W\times L\times t/(W\times (W\times \sin\alpha/2+L/2)\times W\times \sin\alpha/2)$.

\subsection*{Experimental methods}
We performed quasi-static stretching and compression tests of a single kirigami unit cell in a universal testing machine (Instron 5985) equipped with a 300-kN load cell, stretching grippers, and compression plates, which allows us to impose a stretching and compressive displacement with an accuracy of 0.01 mm and to record the force with an accuracy of $0.5\%$ of the reading. We performed a continuous stretching and compressing test in displacement control with a speed of 0.5 mm/s and an out-of-plane compression test with a speed of 0.2 mm/s.  

\begin{figure*}[htb!]
    \centering
\includegraphics[width=0.85\linewidth]{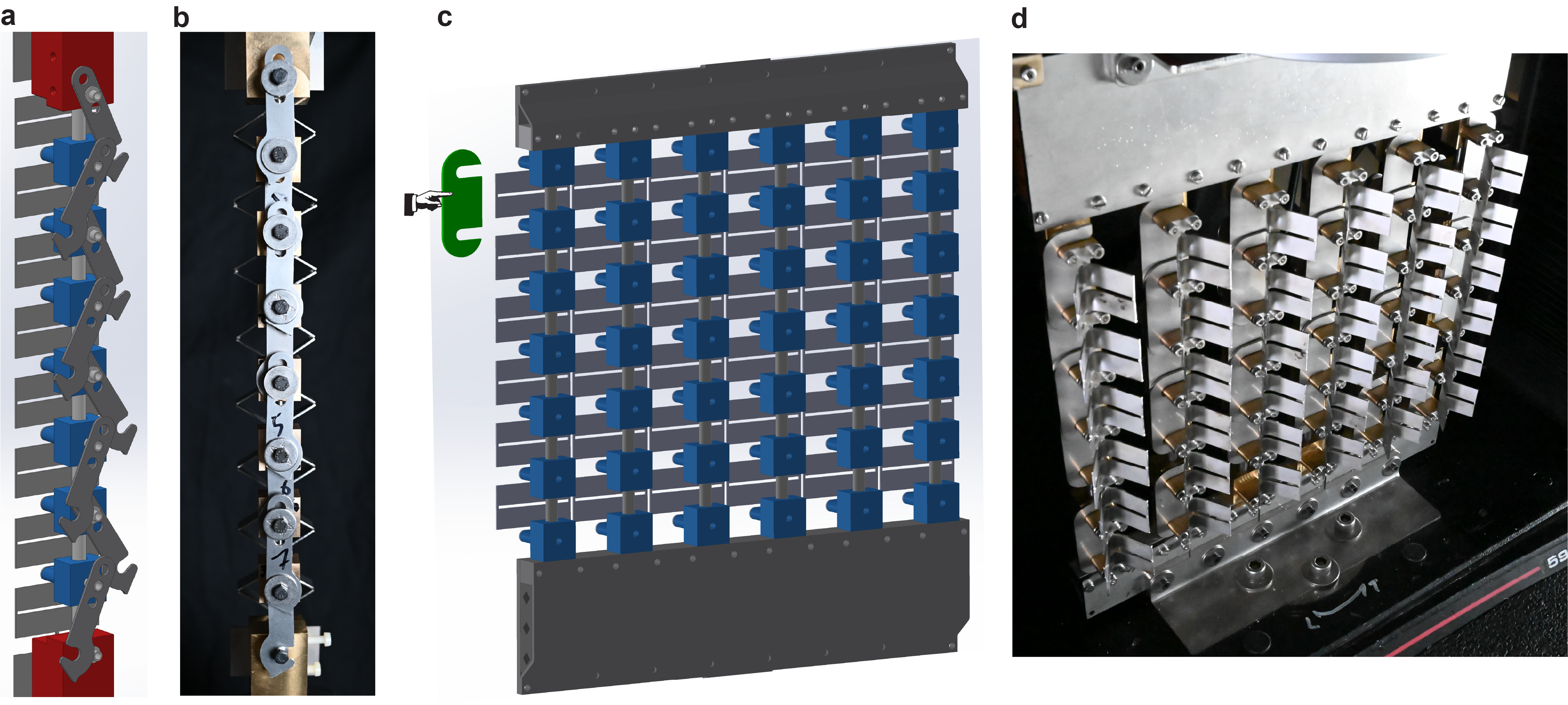}
\caption{\textbf{Setup for sequential self-guided folding of a kirigami strip and a kirigami metamaterial.} \textbf{a and b,} Back view of a setup for a kirigami strip at rest (a) and a snapshot of the setup after folding (b). \textbf{c and d,} Back view of a setup for a kirigami metamaterial at rest (c) and the setup after folding (d). }
    \label{fig:setup}
\end{figure*}

For a kirigami strip and a two-dimensional kirigami under loading cycles, we assemble the kirigami unit cells between two sliding blocks on the same column. The blocks can freely move along the column. We fixed one side of the kirigami and only allowed movement on the other side along the main axial direction. A hook structure is assembled at the back of the sliding block to keep the folding configuration for the rest of the loading cycles (Fig.~\ref{fig:setup}). We designed two types of hook structures: one that self-locks the morphed unit cell under loading back (Fig.~\ref{fig:setup}ab), and the other that requires manual insertion after each loading cycle (Fig.~\ref{fig:setup}cd). 
The quasi-static tests were recorded using a high-resolution camera (Nikon D780 with 105 mm lens, resolution 1080 px $\times$ 1280 px, and frame rate 30 fps). 
\subsection*{Numerical methods}
We use the ``Standard'' solver in the commercial software Abaqus (2020) for the finite-element simulation.

\textit{Model definition}. We analyze the self-folding behavior of a kirigami unit cell and metamaterial with a bilinear elastoplastic model of Young's modulus $E = 200$ GPa, Poisson's ratio $\mu=0.3$, yield stress $\sigma_{\text{y}}$ = 300 MPa, tangent modulus $E_t = 2.3 $ GPa, and an isotropic work hardening behavior. We use 3D stress conditions with eight-node linear brick elements (C3D8R) and construct the mesh to ensure that at least three elements span the thickness direction of the kirigami model and five elements across for all the ligaments.

\textit{Boundary conditions}. For a kirigami unit cell, we fix one side of the unit cell and apply uniaxial stretching on the other side. For the kirigami with multiple unit cells, we constrain all the connected joints between each unit cell and only allow movement along the stretching direction.
% \begin{figure*}[htb!]
%     \centering
% \includegraphics[width=0.8\linewidth]{Figures/A quarter model.png}
% \caption{\textbf{Kirigami unit cell and a quarter of simulation model.}}
%     \label{fig:2}
% \end{figure*}

\textit{Analysis}. We apply quasi-static stretching and compression steps as a single cycle of analysis for a single kirigami unit cell and multiple cycles of analysis for the kirigami strip and 2D kirigami. In the stretching step, we add a small displacement (0.01 mm) to the middle of the unit cell in the out-of-plane direction as the imperfection from the nonlinear buckling analysis.

% \bibliography{Reference}

\clearpage
\setcounter{equation}{0}
\renewcommand{\theequation}{A\arabic{equation}}%
\setcounter{figure}{0}
\renewcommand{\figurename}{\EDF}%
\renewcommand{\figurename}{Fig.}

\begin{appendix}

\input{SI_as_method}

\end{appendix}

\end{document}

%% file: SI_as_method.tex
\clearpage

\begin{center}

{\Large\bfseries
Supplementary Information for\\[0.3cm]
Folding of multishape metamaterials guided by plasticity
\par}

\vspace{0.6cm}

Wenfeng Liu$^{a,b}$,
Jean-Baptiste Chombart$^{a}$,
Bernard Ennis$^{c}$,
Corentin Coulais$^{a,c}$

\vspace{0.4cm}

{\small
$^{a}$Institute of Physics, Universiteit van Amsterdam,
1098 XH Amsterdam, The Netherlands\\[0.2cm]
$^{b}$John A. Paulson School of Engineering and Applied Sciences, Harvard University, Cambridge, Massachusetts 02138, USA\\[0.2cm]
$^{c}$VYTL Works, Amsterdam
}

\end{center}

\vspace{0.8cm}

\renewcommand{\thefigure}{S\arabic{figure}}%
This PDF file includes:
\begin{itemize}
    \item Supplementary text
    \item Figs. S1, S2 and S3
    \item Captions for supplementary movies 1-4
    \item  References for SI reference citations
\end{itemize}

\clearpage

In this supplementary document, we theoretically derive the criteria for elastic buckling and yield buckling of kirigami (sections A and B). In Section C, we investigate the plastic deformations under stretch. Finally, in Section D, we derive the critical condition for folding into \Square\ or \V\ configurations under compression.

\subsection{Elastic buckling analysis}
\begin{figure*}[b!]
    \centering\includegraphics[width=0.95\linewidth]{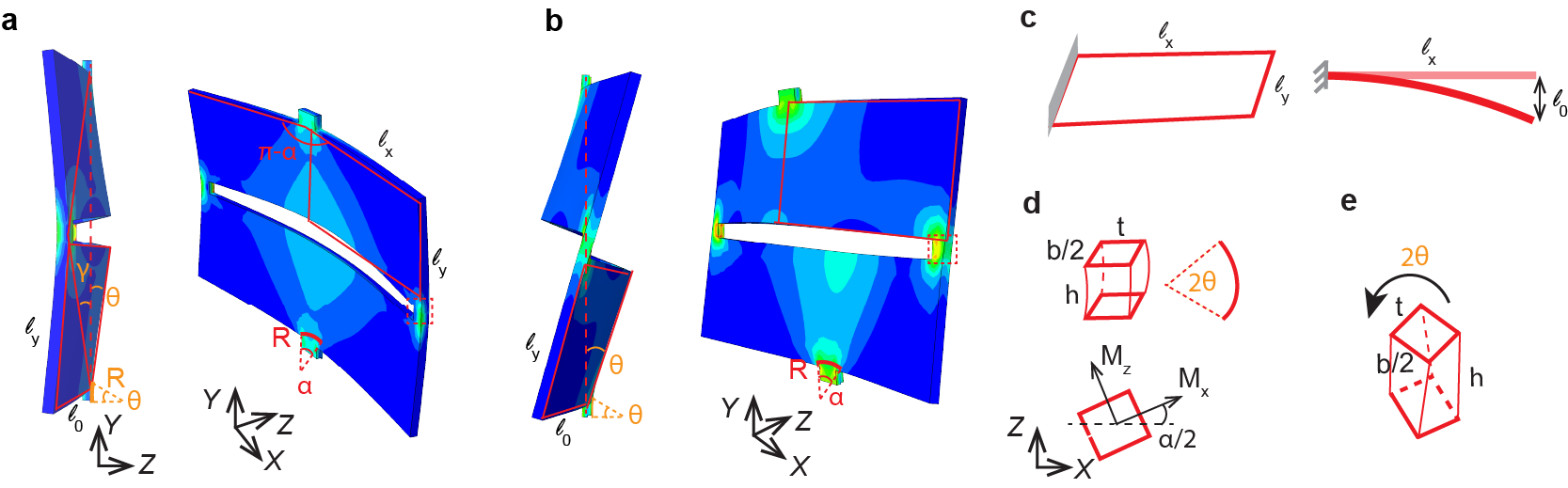}
\caption{\textbf{Buckling analysis.} \textbf{a,} Symmetric buckling mode. \textbf{b,} Antisymmetric buckling mode. \textbf{c,} Sketch of cantilever beam model. \textbf{d,} Sketch of hinge under bending in two directions of symmetric buckling. \textbf{e,} Sketch of hinge under twisting of antisymmetric buckling. }
    \label{fig: buckling analysis}
\end{figure*}

For a unit cell with a parallel pattern of cuts, there are two---symmetric and antisymmetric---possible buckling modes under stretching (Fig.~\ref{fig: buckling analysis}ab). Here, we theoretically analyze the elastic buckling loads of each mode and determine the dominant mode as a function of geometry.

For the symmetric buckling mode, joint angle $\theta$, plate angle $\alpha$, and angle $\gamma$ are coupled, and any two of them can determine the stretching deformation (Fig.~\ref{fig: buckling analysis}a). The relationship between the stretch $\delta_T$ and the projection length of the deflection $\ell_0$ is given by
\begin{equation}
    \delta_T = 2\ell_y\left(\frac{\cos\gamma}{\cos(\gamma+\theta)} - 1\right),\quad \ell_0=\ell_y\tan(\gamma+\theta).
    \label{stretch relation}
\end{equation}

The strain energy of the unit cell comes from two parts: deformed plates and ligaments. We treat each deformed plate as a cantilever beam with a fixed boundary on the left side and a free boundary on the other side (Fig.~\ref{fig: buckling analysis}c). The deflection $\delta(x)$ of the cantilever beam under a concentrated load on the right side can be calculated as $\delta(x) = \ell_0x^2(3\ell_x-x)/2\ell_x^3$, where $\ell_0$ is the deflection at the tip. The internal moment $M(x)$ of the deflected plate can be calculated as $ M(x) = \frac{3EI\ell_0(\ell_x-x)}{\ell_x^3}$, where $EI=E\ell_yt^3/12$ is the bending stiffness of the plate. The bending energy of the plates can be calculated by 
\begin{equation}
     U_p=4\int_{0}^{\ell_x} \frac{M(x)^2}{2EI}dx=\frac{Et^3\ell_y\ell_0^2}{2\ell_x^3}=\frac{Et^3\ell_y^3}{2\ell_x^3}\tan^2(\gamma+\theta).
\end{equation}

For the ligaments at the top and the bottom, the strain energy can be calculated from a deformed tape spring model~\cite{yang2018multistable} as $\mu_{tb} = \frac{1}{2}D(\kappa_t^2+2\nu\kappa_t\kappa_l+\kappa_l^2)$, where $\mu_{tb}$ is the bending energy density, $D=\frac{Et^3}{12(1-\nu^2)}$ is the bending stiffness of a plate with thickness $t$, $\nu$ is the Poisson's ratio, and $\kappa_t$, $\kappa_l$ are the curvatures in the transverse and longitudinal directions (Fig.~\ref{fig: buckling analysis}a orange and red curves of angles $\theta$ and $\alpha$). We assume these two curvatures have the same value $\kappa_t=\kappa_l=1/R$ (Fig.~\ref{fig: buckling analysis}a).
The strain energy of the ligament can be written as $u_{tb} = \mu_{tb}(\alpha R)(\theta R)=D(1+\nu)\alpha \theta$, where $\alpha$ and $\theta$ are corresponding angles of the bent ligament in the transverse and longitudinal directions, respectively. We assume the arc length of angle $\alpha$ is equal to the width of ligament $b$ and obtain $\alpha=b/R$. The curvature $1/R$ equals the bending curvature at the boundary of the plate where $\frac{d^2\delta(x)}{dx^2}=\frac{3\ell_{0}}{\ell_x^2}=\frac{1}{R}$. Thus, the strain energy of the top and the bottom ligaments is given by 
\begin{equation}
    U_{tb} = 2u_{tb}=\frac{Et^3b\ell_y}{2(1-\nu)\ell_x^2}\theta\tan(\gamma+\theta)
\end{equation}
Now, we calculate the bending energy of the middle ligament under two-direction bending (Fig.~\ref{fig: buckling analysis}a and Fig.~\ref{fig: buckling analysis}d). The bending moments in the middle ligament can be written as $M_x=EI_x \kappa\cos\frac{\alpha}{2}$ and $ M_z=EI_z\kappa\sin\frac{\alpha}{2}$, 
where $I_x=\frac{bt^3}{24}$ and $I_z=\frac{tb^3}{96}$ are the second moments of area of the rectangular cross-section of width $b/2$ and height $t$, $\kappa=\frac{2\theta}{h}$ is bending curvature and the angle $\alpha/2$ is the inclination of the bending plane (Fig.~\ref{fig: buckling analysis}d). The bending energy of the ligament can be calculated as
\begin{equation}
    U_m=\int_{0}^{h}\left( \frac{M_x^2}{2EI_x} +\frac{M_z^2}{2EI_z} \right)ds=\frac{Eh(I_x\cos^2\frac{\alpha}{2}+I_z\sin^2\frac{\alpha}{2})}{2}\kappa^2=\frac{Eh(I_x+(I_z-I_x)\sin^2\frac{\alpha}{2})}{2}\kappa^2
    \label{bending energy of hinge}
\end{equation}
For a small deformation, we assume $\sin\frac{\alpha}{2}\approx\frac{\ell_0}{\ell_x}=\frac{\ell_y}{\ell_x}\tan(\gamma+\theta)$. Eq.~\eqref{bending energy of hinge} can be simplified as
$$U_m=\frac{4EI_x}{h}\theta^2+\frac{4E(I_z-I_x)\ell_y^2}{h\ell_x^2}\theta^2\tan^2(\gamma+\theta)$$

Now that we have calculated all the energy terms and work, we are ready to write the Lagrangian formulation $\mathcal{L}(\gamma,\theta)$ of the symmetric mode as 
\begin{equation}
    \mathcal{L}(\gamma,\theta)=U_p+U_{tb}+U_m+F\left(\delta_T - 2\ell_y\left(\frac{\cos\gamma}{\cos(\gamma+\theta)} - 1\right)\right).
\end{equation}
We then minimize the potential energy with respect to $\gamma$ and $\theta$ and let $\frac{\partial\mathcal{L}}{\partial\gamma}=0$, $\frac{\partial\mathcal{L}}{\partial\theta}=0$, and $\frac{\partial^2\mathcal{L}}{\partial\gamma\partial\theta}=0$, simultaneously. We use a first-order Taylor expansion for small values of $\gamma$ and $\theta$ ($\cos\theta=1$, $\sin\theta=\theta$) to calculate the critical force $F^s_{cr}$ of symmetric buckling as
\begin{equation}\label{eq:symmtric buckling load}
    \frac{F^s_{cr}\ell_y}{Et^3}=\frac{1}{2}\left[\frac{\ell_y^3}{\ell_x^3}+\frac{b\ell_y}{2(1-\nu)\ell_x^2}\right]
\end{equation}

For the antisymmetric buckling (Fig.~\ref{fig: buckling analysis}b), the stretch value $\delta_T$ can be determined by a single angle $\theta$
\begin{equation}
    \delta_T = 2\ell_y\left(\frac{1}{\cos\theta} - 1\right).
\end{equation}
The bending energy of the plates $U_b$ and the top and the bottom ligaments $U_tb$ have the same form as the symmetric buckling mode
\begin{equation}
     U_p=\frac{Et^3\ell_y^3}{2\ell_x}\tan^2(\theta)
\end{equation}

\begin{equation}
    U_{tb} = \frac{Et^3\ell_y}{2(1-\nu)\ell_x^2}\theta\tan(\theta)
\end{equation}
The middle ligaments of the antisymmetric buckling unit cell are under a torsional deformation (Fig.~\ref{fig: buckling analysis}e). The relationship between the torque $T$ and torsional angle $2\theta$ can be defined as $T=\frac{2GJ\theta}{h}$, where $G=\frac{E}{2(1-\nu)}$ is the shear modulus, and $J$ is the second polar moment of the cross-section area. For a rectangular cross-section of width $b/2$ and depth $t$, so that $ J  
\approx \frac{1}{24} tb^3$ when $b \ll 2t$ and $
J \approx \frac{1}{6} bt^3$ when $b \gg 2t$.
The torsional energy of the ligaments can be calculated as   
\begin{equation}
     U_m=2\int_{0}^{h} \frac{T(x)^2}{2GJ}dx=\frac{EJ}{h(1-\nu)}\theta^2.
\end{equation}
With all the energy terms, we can write the Lagrangian $\mathcal{L}(\theta)$ for the antisymmetric buckling mode as
\begin{equation}
     \mathcal{L}(\theta)=\frac{Et^3\ell_y^3}{2\ell_x}\tan^2(\theta) +\frac{Et^3\ell_y}{2(1-\nu)\ell^2}\theta\tan(\theta)+\frac{EJ}{h(1-\nu)}\theta^2+F\left(\delta_T-2\ell_y\left(\frac{1}{\cos\theta} - 1\right)\right)
\end{equation}
With $\frac{\partial\mathcal{L}}{\partial\theta}=0$, we obtain the critical force of antisymmetric buckling as
\begin{equation}\label{eq: antisymmetric buckling load}
    \frac{F^a_{cr}\ell_y}{Et^3}=\frac{1}{2}\left[\frac{\ell_y^3}{\ell_x^3}+\frac{b\ell_y}{2(1-\nu)\ell_x^2}+\frac{8J}{ht^3}\right].
\end{equation}

We combine Eq.~\eqref{eq:symmtric buckling load} and Eq.~\eqref{eq: antisymmetric buckling load} and obtain the relation between $\tilde{F}^s_{cr}$ and $\tilde{F}^a_{cr}$ as
\begin{equation}
\boxed{
\tilde{F}^a_{cr}=\tilde{F}^s_{cr}+\frac{4J}{ht^3}}
\end{equation}
which is Eq. (4) in the Methods. We can find that the symmetric buckling load is smaller than the antisymmetric buckling load. Only in the case of $b\ll 2t$ ($\frac{8J}{ht^3}\approx\frac{b^3}{3ht^2}\approx0 $ and $F^a_{cr}\approx F^s_{cr}$) may the antisymmetric buckling mode occur (as shown by our numerical simulations in Fig. 5 of the Methods). 

\subsection{Yield buckling regime}
 Now, let us consider the buckling behavior of elastoplastic kirigami. When the thickness of the kirigami is small enough, or the width of the kirigami is large enough (Eq.~\eqref{eq:symmtric buckling load}), the kirigami buckles before the material yields and plastic deformation only occurs in postbuckling due to overstretch---this is elastic buckling. As the thickness of the kirigami increases or its width decreases, the kirigami buckles after the hinges plastify---this is yield buckling. Yield buckling arises from a highly asymmetric stress distribution at the hinges along the loading path and results in negative stiffness at the onset of buckling~\cite{liu2024harnessing}. Thus, we can write the regime of yield buckling as $F_y<F_{cr}$, where $F_y$ is the yield load of the hinges and $F_{cr}$ is the elastic critical load. When we use the symmetric buckling load $F_{cr}=F^s_{cr}$ in Eq.~\eqref{eq:symmtric buckling load} as the elastic critical load and $F_{y}=\sigma_y bt$ as the yield load of the hinges, we can obtain the yield buckling regime as 
\begin{equation}\label{eq:yield buckling-2}
\boxed{
    \frac{\sigma_y}{E}<
    \left[\frac{\ell_y^2}{2b\ell_x}+\frac{1}{4(1-\nu)}\right]\left(\frac{t}{\ell_x}\right)^2}  
\end{equation}
which is Eq. (5) in the Methods.

\subsection{Plastic deformation in postbuckling of stretch}
\begin{figure*}[b!]
    \centering                                                   \includegraphics[width=0.4\linewidth]{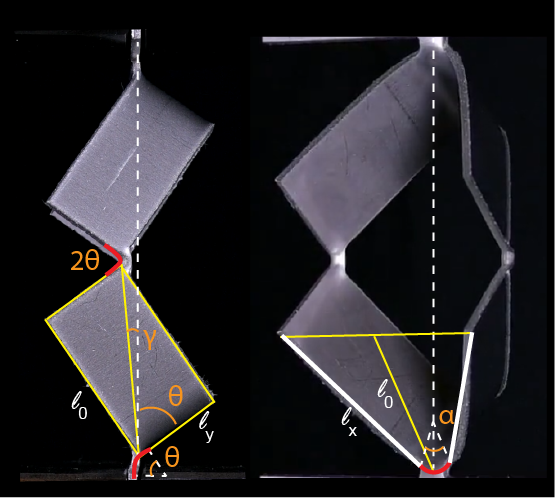}
\caption{\textbf{Stretching deformation analysis.} \textbf{a and b,} Snapshots of the symmetric (a) and antisymmetric (b) buckling unit cells at the side view and oblique view. \textbf{c,} Sketches of a deformed plate of the unit cell with a fixed boundary condition at one side and a concentrated force condition at the other side. }
    \label{fig: postbuckling analysis}
\end{figure*}

For a kirigami that buckles in the regime of yield buckling, the hinges act as plastic hinges in postbuckling. Thus, we can simplify the analysis by assuming that the deformed kirigami consists of four rigid plates connected by plastic hinges. There are three angles of deformation in the stretching, $\gamma$, $\theta$, and $\alpha$ (Fig.~\ref{fig: buckling analysis} and Fig.~\ref{fig: postbuckling analysis}), and any two of them can determine the deformation. Now we eliminate $\gamma$ in Eq.~\eqref{stretch relation} and use the hinge rotating angle $\theta$ and the plates' bending angle $\alpha$ to describe the deformation. We first simplify the projection length of the deflection $\ell_0$ in Eq.~\eqref{stretch relation} as $ \ell_0\approx \ell_x\sin\frac{\alpha}{2}$, and $\tan(\gamma+\theta)=\frac{\ell_0}{\ell_y}=\frac{\ell_x}{ \ell_y}\sin{\frac{\alpha}{2}}$. Thus, the stretch value $\delta_{T}$ in Eq.~\eqref{stretch relation} can be written as
\begin{equation}\label{eq:strech relation in postbuckling}
    \delta_{T}\left(\theta, \alpha\right)=2 \ell_{y}\left(\cos \theta-1\right)+2 \ell_{x}\sin\frac{\alpha}{2}\sin \theta,
\end{equation}
For the whole stretching process we can write the Lagrangian $\mathcal{L}(\theta, \alpha)$ as
\begin{equation}
    \mathcal{L_T}( \theta, \alpha)=2 \int_{0}^{\theta} M\left(\theta^{'}\right) d \theta^{'}+ \int_{0}^{2\theta} M\left(\theta^{'}\right) d \theta^{'} + 2 \int_{0}^{\alpha} M\left(\alpha^{'}\right) d \alpha^{'}+F\left(\delta_T-\left(2 \ell_{y}\left(\cos \theta-1\right)+2 \ell_{x}\sin\frac{\alpha}{2} \sin \theta\right)\right),
\end{equation}
where $M(\theta)$ and $M(\alpha)$ are the moments at the hinge joint and the plate joint, $F$ is a Lagrange multiplier that corresponds to the axial force. With  $\frac{\partial\mathcal{L}}{\partial\theta}=0$ and $\frac{\partial\mathcal{L}}{\partial\alpha}=0$, we obtain the moment balance of the hinge joint and the plate joint as
\begin{equation}\label{eq:momment balance}
\begin{cases}
\frac{\partial\mathcal{L}}{\partial\theta}=2M(\theta)+2M(2\theta)+2F\left(\ell_y\sin\theta- \ell_{x}\sin\frac{\alpha}{2}\cos\theta\right)=0,\\
\large\frac{\partial\mathcal{L}}{\partial\alpha}=2M(\alpha)-F\ell_x\sin\theta\left. \cos\frac{\alpha}{2}=0\right.
\end{cases}
\end{equation}
We can eliminate $F$ in Eq.~\eqref{eq:momment balance} and obtain an implicit kinematic relation between $\theta$ and $\alpha$:
\begin{equation}
(M(\theta)+M(2\theta))\sin\theta\cos\frac{\alpha}{2} +\,2M(\alpha)\Big(\frac{\ell_y}{\ell_x}\sin\theta-\sin \frac{\alpha}{2}\cos\theta\Big)=0,
\end{equation}

To explicitly solve $\theta$ and $\alpha$ as a function of the stretch $\delta_T$, here, we combine the geometric constraint in Eq.~\eqref{eq:strech relation in postbuckling} and use Taylor expansion for $\cos\theta-1 \approx-\frac{\theta^2}{2}$, $\sin\frac{\alpha}{2}\approx \frac{\alpha}{2}$, $\sin\theta \approx \theta$, $\cos\frac{\alpha}{2}\approx1$, and $\cos\theta \approx1$. We simplify the moments $M(\theta)$ and $M(\alpha)$ at the hinge joint and the plate joint with the rigid-plastic model, which leads to $M(\theta)=M(2\theta)=\frac{\sigma_ybt^2}{4}$ and $
M(\alpha)=
\frac{\sigma_y\ell_yt^2}{4}.$ Thus, Eqs.~\eqref{eq:strech relation in postbuckling} and~\eqref{eq:momment balance} can be simplified as
\begin{equation}
    \begin{cases}
\delta_T=-\ell_{y}\theta^{2}+ \ell_{x}\alpha\theta, \\
2\frac{b}{\ell_y}\theta+2\frac{\ell_y}{\ell_x}\theta-\alpha=0,\\
\end{cases}
\end{equation}

\begin{figure*}[htb!]
    \centering                                                   \includegraphics[width=0.7\linewidth]{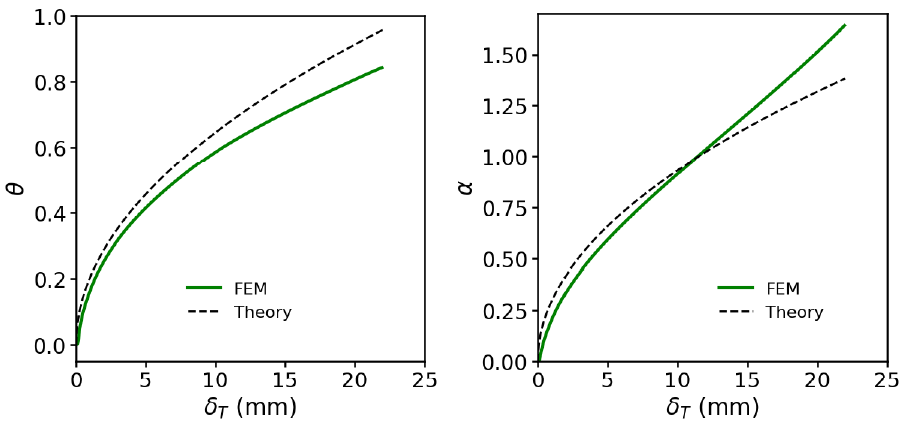}
\caption{\textbf{Finite-element results and predictions of the joint angle $\theta$ and the plate angle $\alpha$ vs. the stretching value $\delta_T$ in a kirigami of $\ell_x=27$ mm, $\ell_y=14$ mm, and $b=2.5$ mm.}}
    \label{fig: stretch FEM and prediction}
\end{figure*}

We can then explicitly solve $\theta$ and $\alpha$ as a relation of $\delta_T$:
\begin{equation}\label{eq: angle change under stretching}
\boxed{
\theta = 
\left(\frac{\ell_y}{\ell_y^2+2b\ell_x}\right)^{\!1/2} \delta_T^{1/2},
\qquad
\alpha = \frac{2(\ell_y^2+b\ell_x)}{\ell_x\ell_y^{1/2}\left(\ell_y^2+2b\ell_x\right)^{\!1/2}} \delta_T^{1/2},}    
\end{equation}
which is Eq. (1) in the Main Text. We then conduct a finite-element simulation in a kirigami and find that our simplified model (Eq.~\eqref{eq: angle change under stretching}) can be used to describe the deformed angle under stretching (Fig.~\ref{fig: stretch FEM and prediction}). Next, we will use these stretching angles as a reference for the analysis of compression.  

\subsection{Folding configurations under compression.}

The stretching angles in Eq.~\eqref{eq: angle change under stretching} become $\theta_p$ and $\alpha_p$ after unloading due to plasticity, which we will use as the reference configuration for the analysis of compression. The compression displacement is
\begin{equation}
    \delta_C = 2\ell_y\left(\frac{\cos\gamma_{p}}{\cos(\gamma_{p}+\theta_{p})} - \frac{\cos(\gamma_{p}+\Delta \gamma)}{\cos(\gamma_{p}+\theta_{p}+\Delta \gamma+\Delta \theta)}\right),
\end{equation}
From the geometric constraints $\tan(\gamma_p+\theta_p) \approx \frac{\ell_x}{ \ell_y}\sin{\frac{\alpha_p}{2}}$ and $ \tan(\gamma_p+\theta_p+\Delta \gamma+\Delta \theta) \approx \frac{\ell_x}{ \ell_y}\sin{\frac{\alpha_p+\Delta \alpha}{2}}$, we can eliminate $\gamma_p$ and $\Delta\gamma$ in $\delta_C$, and then Taylor expand it to the first order of $\Delta \theta$ and $\Delta\alpha$ as
\begin{equation}\label{eq:compressing constraint}
    \delta_C(\Delta\theta, \Delta\alpha) \approx 2\ell_y\left(
        \sin\theta_p
        - \frac{\ell_x}{\ell_y}\sin\frac{\alpha_p}{2}\cos\theta_p
    \right)\Delta\theta
    - \ell_x \sin\theta_p \cos\frac{\alpha_p}{2}\,\Delta\alpha=C\Delta\theta-D\Delta\alpha,
\end{equation}
where $C=2\ell_y\left(
        \sin\theta_p
        - \frac{\ell_x}{\ell_y}\sin\frac{\alpha_p}{2}\cos\theta_p
    \right)$ and $D= \ell_x \sin\theta_p \cos\frac{\alpha_p}{2}\,$.  
Following the same procedure as for stretching, we can write the Lagrangian $\mathcal{L_C}(\Delta\theta, \Delta\alpha)$ for compression as,
\begin{equation}
    \mathcal{L_C}(\Delta\theta, \Delta\alpha)=2 \int_{0}^{\Delta\theta} M\left(\theta^{\prime}\right) d \theta^{\prime}+ \int_{0}^{2\Delta\theta} M\left(\theta^{\prime}\right) d \theta^{\prime} + 2 \int_{0}^{\Delta\alpha} M\left(\alpha^{\prime}\right) d \alpha^{\prime}+F\left(\delta_C-\delta_C(\Delta\theta, \Delta\alpha)\right)
\end{equation}
From $\frac{\partial\mathcal{L_C}}{\partial\Delta\theta}=0$ and $\frac{\partial\mathcal{L_C}}{\partial\Delta\alpha}=0$, we obtain the relations of torque balance during compression as
\begin{equation} \label{eq:compressing moment balance}
    \begin{cases}
        \frac{\partial\mathcal{L_C}}{\partial\Delta\theta}=2M(\Delta\theta)+2M(2\Delta\theta)-FC=0\\
        \large\frac{\partial\mathcal{L_C}}{\partial\Delta\alpha}=2M(\Delta\alpha)+FD=0,
    \end{cases}
\end{equation}

where $\frac{M(\Delta\theta)}{M(\Delta\alpha)}=\frac{b}{\ell_y}$. By eliminating $F$ in Eq.~\eqref{eq:compressing moment balance}, we can obtain 
\begin{equation}\label{eq: angle relation}
    \frac{3b}{\ell_y}D\Delta\theta+C\Delta\alpha=0.
\end{equation}
% \begin{equation}
% \begin{cases}
%     3k_{\theta}\Delta\theta+\frac{C}{D}k_{\alpha}\Delta\alpha+2p+\frac{C}{D}q=0\\
%     \delta_C=C\Delta\theta-D\Delta\alpha
% \end{cases} 
% \end{equation}
From Eq.~\eqref{eq:compressing constraint} and Eq.~\eqref{eq: angle relation}, we can explicitly solve $\Delta\theta$ and $\Delta\alpha$ as
\begin{equation}\label{eq:angle relations in compression}
    \Delta\theta = \frac{C\ell_y\,\delta_C}{C^2\ell_y + 3bD^2}
, \quad
\Delta\alpha = -\frac{3bD\,\delta_C}{C^2\ell_y + 3bD^2}.
\end{equation}
We then consider the critical condition for selecting the \Square\ and \V\ configurations under compression. When the hinge angle $\theta$ increases with the compression $\delta_C$, it folds into the \Square\ shape, and otherwise it gets into the \V\ shape. Thus, we can obtain the critical condition of folding as $\frac{d \theta}{d \delta_C}=\frac{d \Delta\theta}{d \delta_C}=0$. By Substituting into Eq.~\eqref{eq:angle relations in compression}, we have
% \begin{equation}
%     \sin\theta_p
%         - \frac{\ell_x}{\ell_y}\sin\frac{\alpha_p}{2}\cos\theta_p=0
% \end{equation}
\begin{equation}\label{eq:critical condition}
   \tan\theta_p =\tilde{\ell}\sin\frac{\alpha_p}{2},
\end{equation} 
where $\theta_p \approx \left(\frac{2}{1+2\tilde{b}\tilde{\ell}}\tilde{\delta}\right)^{1/2}$, $
\alpha_p \approx  \left(\frac{8(1+\tilde{b}\tilde{\ell})^2}{\tilde{\ell}^2(1+2\tilde{b}\tilde{\ell})}\tilde{\delta}\right)^{1/2}$, $\tilde{\delta} = \frac{\delta_T}{2\ell_y}$, $
\tilde{b} = \frac{b}{\ell_y}$, and $ 
\tilde{\ell}= \frac{\ell_x}{\ell_y}.$ 
Taylor-expanding $\tan\theta_p$ and $\sin\frac{\alpha_p}{2}$ into third order of $\theta_p$ and $\alpha_p$: $\tan\theta_p\approx \theta_p+\frac{\theta_p^3}{3}$ and $\sin\frac{\alpha_p}{2}\approx\frac{\alpha_p}{2}-\frac{\alpha_p^3}{48}$ , the critical condition Eq.~\eqref{eq:critical condition} can be expressed as
\begin{equation}
 \boxed{
\tilde{\delta} \; = \; 
\frac{3\,\tilde{b}\,\tilde{\ell}^3\left(1 + 2\tilde{b}\tilde{\ell}\right)}
     {\,2\tilde{\ell}^2 + \left(1 + \tilde{b}\tilde{\ell}\right)^3\,}
     }   
\end{equation}
which is Eq. (2) in the Main Text.

\clearpage

\subsection{Supplementary Movies}
Description of supplementary movies:\\

Supplementary Movie 1: 
\textbf{Self-guided folding by plasticity
.}
We introduce a kirigami unit cell with a parallel pattern of cuts made from a 316L stainless steel plate. Such kirigami buckles symmetrically under stretching and folds into two distinct three-dimensional structures under compression. The selection of the two self-guided folding structures depends on the stretching stroke. \\

Supplementary Movie 2: \textbf{Additional self-guided folding shapes.} We demonstrate that a kirigami unit with an antisymmetric imperfection can fold into three distinct three-dimensional structures. We then extend kirigami patterns of cuts by varying the position and the number of the hinges, which can fold into various shapes.\\

Supplementary Movie 3: \textbf{Mechanical performance under out-of-plane compression.}
We conduct out-of-plane compression on the three distinct folding structures. We plot Ashby maps based on their stiffness, strength, and specific energy absorption under different values of stretch.  \\

Supplementary Movie 4: \textbf{Shape morphing kirigami programmed by loading history
.} 
We explore a regime of yield buckling in a kirigami strip of eight unit cells, which enables sequential buckling under stretching. We then demonstrate that such a kirigami strip can sequentially fold into a diamond strip under a series of stretching-compression loading cycles. We eventually demonstrate that a $6 \times 6$ kirigami metamaterial consisting of two types of unit cells can morph into three distinct patterns under three different periodic loading histories.

% \bibliography{Reference }

\setcounter{equation}{0}
\renewcommand{\theequation}{S\arabic{equation}}%
\setcounter{figure}{0}
\renewcommand{\figurename}{Supplementry Fig.}